\documentclass[preprint,12pt]{elsarticle}
\usepackage[small,it]{caption}
\usepackage{cancel}
\usepackage{layout}
\usepackage{amsmath,amssymb,amsfonts,mathrsfs,amscd,euscript,enumerate,calc,verbatim}
\usepackage{graphicx}
\usepackage{subcaption}
\usepackage{latexsym}
\usepackage{float}
\usepackage{makecell}
\usepackage{array}
\usepackage[usestackEOL]{stackengine}
\graphicspath{{figures/}}
\usepackage{algorithm}
\usepackage[noend]{algorithmic}
\usepackage{color}
\usepackage{todonotes}

\usepackage{tikz}
\usepackage{pgfplots}
\usetikzlibrary{automata}
\usepackage{caption}
\usepackage{amsmath,amssymb,amsfonts}
\usepackage{graphicx}
\usetikzlibrary{decorations.pathreplacing,calligraphy,  arrows,positioning,fit,calc}
\tikzstyle{block} = [rectangle, minimum width=5em, minimum height=0.4em, draw=black, text width=8em,
text centered]
\tikzstyle{rec} = [rectangle, minimum width=22em, minimum height=0.4em, draw=black, text width=15em,
text centered]
\tikzstyle{sqr} = [square/.style={regular polygon,regular polygon sides=4}]
\tikzstyle{line} = [draw, -latex']

\tikzstyle{rec1} = [rectangle, minimum width=19em, minimum height=0.4em, draw=black, text width=19em,
text centered,dashed]
\tikzstyle{sqr1} = [rectangle, minimum width=1.8em, minimum height=1.8em, draw=black, text width=1.6em,text centered]
\tikzstyle{sqrb} = [rectangle, minimum width=7em, minimum height=1.8em, draw=black, text width=7em,
text centered,thick]
\tikzstyle{rec3} = [rectangle, minimum width=23em, minimum height=0.4em, draw=black, text width=23em,
text centered,dashed]
\tikzstyle{sqr3} = [rectangle, minimum width=1.8em, minimum height=1.8em, draw=black, text width=1.6em,
text centered]
\tikzstyle{sqrb3} = [rectangle, minimum width=1.8em, minimum height=1.8em, draw=black, text width=1.6em,
text centered,thick]

\newtheorem{theorem}{Theorem}[section]

\newtheorem{definition}[theorem]{Definition}

\newtheorem{remark}[theorem]{Remark}

\def\E{\texttt{E}}

\def\kM{\texttt{kXor}}
\def\eL{\texttt{eLeft}}
\def\eR{\texttt{eRight}}
\def\dL{\texttt{dLeft}}
\def\dR{\texttt{dRight}}
\def\RO{$\texttt{Rnd}_{\texttt{odd}}$}
\def\RE{$\texttt{Rnd}_{\texttt{even}}$}

\def\l{{\mathsf{l}}}

\usepackage{url}

\usepackage{xcolor}
 \usepackage[english]{babel}
\makeatletter
\@ifpackageloaded{babel}
  {\newenvironment{nohyphens}
     {\par\sloppy\exhyphenpenalty=\@M
      \@ifundefined{l@nohyphenation}
        {\language=\@cclv}
        {\hyphenrules{nohyphenation}}%
     }
     {\par
      \@ifundefined{l@nohyphenation}
        {}
        {\endhyphenrules}%
     }
  }
  {\newenvironment{nohyphens}
     {\par\sloppy\exhyphenpenalty=\@M
      \@ifundefined{l@nohyphenation}
        {\language=\@cclv}
        {\language=\l@nohyphenation}%
     }
     {\par}
  }
\makeatother 

\begin{document}

\sloppy
	\begin{frontmatter}
	\title{INRU: A Quasigroup Based Lightweight Block Cipher} 
		
		\author[1]{Sharwan K. Tiwari\corref{cor1}}
		\ead{shrawant@gmail.com}
		\cortext[cor1]{Corresponding author}
		\author[1]{Ambrish Awasthi}
		\ead{ambrishawasthi@yahoo.com}
		\author[1]{Sucheta Chakrabarti}
		\ead{suchetadrdo@hotmail.com}
		\author[1]{Sudha Yadav}
		\ead{sudhayadav2611@gmail.com}
		
		\address[1]{Scientific Analysis Group, Defence Research $\&$ Development Organization, Metcalfe House, Delhi-110054, India}

		\begin{abstract}
\begin{nohyphens}
In this paper, we propose a quasigroup based block cipher design. 
The round functions of the encryption and decryption algorithms use quasigroup based string transformations. We show the robustness of the design against the standard differential, linear and algebraic cryptanalytic attacks. We also provide detailed statistical analysis using NIST test suite in CBC, CFB, OFB, and CTR modes of operation. We compare the statistical experimental results with the AES-128 in the same setup and conclude that the randomizing ability of our algorithm is equivalent to that of AES-128.
\end{nohyphens}  
\end{abstract}

	\end{frontmatter}

\section{Introduction}
\begin{nohyphens}
Quasigroup based cryptography is rapidly evolving and gaining the attention of the cryptographic community for the last two decades \cite{Belyavskaya1994, viktor2017, smith2007}. Quasigroups have found many applications in the various cryptographic primitives and algorithms\cite{Mileva2010}. They are used in the construction of  Sboxes, pseudorandom number generators, and hash functions\cite{10.1007/11502760_11,Gligoroski2009EdonRAI}. They are also used  in message authentication, zero-knowledge protocol, and in the designing of the fast stream and block ciphers \cite{Gligoroski2008APK, Battey2013AnEQ, Dnes1992ANA,  6394451}. Several quasigroup based cryptographic schemes viz. Edon-80, Edon-R, and GAGE $\&$ InGAGE, etc. \cite{Gligoroski2008TheSC, gage} participated in the various famous cryptographic competitions and had been analyzed by the community. 

The non-associative and non-commutative properties of the quasigroup structure make them useful for the cryptographic purpose because many well-known standard cryptanalytic techniques to attack quasigroup based cryptographic designs cannot be applied straightforward. 

The Quasigroup structure defines a combinatorial structure known as Latin square, and conversely. They are well-studied algebraic and combinatorial structures, see in \cite{Belyavskaya1989, Belyavskaya1992, Keedwell2015, Denes1991, kepka1978, kepka1971}. Quasigroup string transformations are one of the fundamental building blocks in the design of quasigroup based cryptographic algorithms. The security of these transformations heavily depends on a proper choice of quasigroups \cite{Artamonov2016, Artamonov2020, smile1997, Markovski1999}. In particular, the choice of quasigroups affects the rate of growth of the period and randomness of their output \cite{Dimitrova04}. Of course, a significant period and randomness provide good security to the cryptographic algorithms.

A suitable class of quasigroups for the cryptographic purpose should consist of non-affine $\&$ simple quasigroups and they should not possess any proper subquasigroup. Since string transformations over quasigroups belonging to this class provide high non-linearity in the cryptographic primitives and introduce necessary confusion in the block cipher algorithms.  Furthermore, these properties in the quasigroup also ensure that the search space of the brute-force attack cannot be reduced. The simple and non-affine quasigroups are polynomially complete algebraic structures and the problem of solving a system of equations over such algebraic structures is NP-complete \cite{Horv2008}.

We propose a block cipher design based on quasigroup string transformations. The string transformation based cryptographic algorithms use different functions for encryption and decryption. These functions use string transformations over conjugate quasigroups. The conjugates of suitable finite quasigroups are suitable \cite{normal}. The security strength of such cryptographic schemes needs to be evaluated corresponding to both functions. We analyze the design rigorously against the standard linear and differential cryptanalytic attacks and show that these well-known techniques do not apply straightforward. We have also generated an algebraic system of polynomial equations over $\mathbb{F}_2$ of the design by using a Boolean representation of the quasigroup binary operation. We perform some practical experiments of solving  algebraic systems of the reduced round of the design using Gr\"obner bases techniques. The complete algebraic system is too complex to be solved using only algebraic techniques.

The organization of the paper is as follows. In Section \ref{s1}, we revise some definitions and properties of the quasigroups used in this paper. Then, in Section \ref{s2}, we present our block cipher design.  In section \ref{s3}, we analyze our design using standard linear, differential, and algebraic cryptanalytic techniques and show the robustness of the design against these attacks. Further, in Section \ref{s4}, we also test the randomizing ability of our algorithm using NIST statistical test suite\cite{10.5555/2206233}. We compare the statistical experimental results with AES-128 in the same setup and conclude that the randomization obtained from both  algorithms is equivalent. We conclude the work in Section \ref{s5}.
\end{nohyphens}
%
%
%
%
%

\section{Preliminaries}\label{s1}
In this section, we introduce some of the terminology used in this paper.\\
 A \emph{quasigroup} is a set $Q$ with a binary operation of multiplication
  $x \ast y$ such that for all $a,b\in Q$ the equations 
 $
 a\ast x=b,\quad y\ast a=b
 $
 have unique solutions 
 $
 x=a\backslash_\ast b ,\quad y=b /_\ast a.
 $
 A binary operation $\ast$ of quasigroup provides two other binary operations $\backslash_\ast$ and $/_\ast $ which are defined as 
\begin{eqnarray*} \label{eq1}
x\backslash_\ast(x \ast y)=y=x\ast (x\backslash_\ast y),~
(x\ast y )/_\ast y=x=  (x /_\ast y)\ast y,~
~\forall x,y\in Q.
\end{eqnarray*}  
Each quasigroup $(Q=\{x_1,\ldots,x_n\},\ast)$ can be given by a Latin square $L$  of size $n^2$
\begin{eqnarray} \label{eqls1}
  \begin{array}{ c|ccc}
  \ast & x_1 & \hdots & x_n\\
  \hline
 x_1 &   a_{11}  & \hdots & a_{1n} \\ 
\vdots &  \hdots & \hdots  &  \hdots\\
 x_n &   a_{n1}   & \hdots & a_{nn} \\
  \end{array}
  \end{eqnarray}
Each entry $a_{ij}$ stands for the product $x_i\ast x_j$ in the quasigroup $(Q,\ast)$. 
A quasigroup $(Q,\ast)$ is  \emph{affine} if $Q$   is  equipped with a structure of additive abelian group $(Q,+)$ such that $\forall$ $x,y\in Q$,  
$$x\ast y=\alpha (x)+\beta (y)+c$$
where $c\in Q$ and $\alpha,\beta$ are automorphisms of $(Q,+)$ .
A quasigroup is \emph{simple} if  it has only trivial congruence relations. A finite quasigroup is said to be \emph{polynomially complete} if it is simple and non-affine.
\begin{definition}[\texttt{e}-transformation]
Let $(Q,\ast)$ be a quasigroup and $Q^r=\{a_0,a_1,\ldots,a_{r-1} \mid a_i \in Q, r \geq 2 \}$. Fix an element $l \in Q$, commonly known as leader, define
\begin{eqnarray*}
\texttt{e}_{{(Q,\ast),l}} : Q^r \rightarrow Q^r, \quad \quad a_0,\ldots,a_{r-1}\mapsto b_0,\ldots,b_{r-1}\\
b_0 = l \ast a_0, \quad b_i = b_{i-1} \ast a_i, \quad i = 1,\ldots,r-1
\end{eqnarray*}
The map $\texttt{e}_{{(Q,\ast),l}}$ is generally known as {\emph (left) \texttt{e}-transformation}. We denote $\texttt{e}_{{(Q,\ast),l}}$ by $\texttt{e}_{l}$, when the quasigroup is clear in the context.
\end{definition}
\begin{definition}[\texttt{d}-transformation] 
Define another elementary transformation $\texttt{d}$ on $(Q,\backslash_{\ast})$, corresponding to $(Q,\ast)$, with leader $l \in Q$ given by 
\begin{eqnarray*}
\texttt{d}_{(Q,\backslash_{\ast}),l} :Q^r \rightarrow Q^r, \quad \quad a_0,\ldots,a_{r-1}\mapsto b_0,\ldots,b_{r-1}\\
b_0 = l \backslash_{\ast} a_0, \quad b_i = a_{i-1} \backslash_\ast a_i, \quad i = 1,\ldots,r-1
\end{eqnarray*}
It is commonly known as {\emph(left) \texttt{d}-transformation}.  
\end{definition}
These transformations are also commonly known as {Quasigroup (elementary) String Transformations}.
Let $(Q,\ast,/_{\ast},\backslash_{\ast})$ be a finite quasigroup, then for each 
$a \in Q^r $ and any fix $l \in Q$, the  elementary transformations 
$\texttt{e}_l\text{ and }\texttt{d}_l$, defined above,  are mutually inverse maps, i.e.
  $$\texttt{d}_{(Q,\backslash_{\ast}),l}\circ \texttt{e}_{{(Q,\ast),l}}(a)~= a~=~\texttt{e}_{{(Q,\ast),l}}\circ \texttt{d}_{(Q,\backslash_{\ast}),l}(a)$$

Let  $\ast$ and $\ast'$ be  two binary operations on a set $Q$ and consider  $l,l'\in Q$. Then, the composite (left) transformations, denoted by  $\E$  $\&$ $\texttt{D}$ are defined as follows: 
$$\hspace*{16mm} \E_{l,l'}=e_{l,(Q,\ast)}~\circ ~ e_{l',(Q,\ast')},\quad \texttt{D}_{l',l}=d_{l',(Q,\backslash_{\ast'})}~\circ~ d_{l,(Q,\backslash_{\ast})}
$$
The composition can be defined for any finite number of string transformations over quasigroups.  
It is to be noted that for the composite string transformation, all the leaders and quasigroup operations need  not  be distinct. These transformations play an important role to increase the security of quasigroup based cryptographic algorithms by using more than one quasigroups and more than one round with different leaders. 
In the next section we propose a new block cipher design based on quasigroup using string transformations.
\section{Block cipher design based on Quasigroup string transformations}\label{s2}
Our design of block cipher consists of $16$ rounds. It operates on a block size of $64$-bit and requires $128$-bit long key. The round function is mainly defined by elementary  string transformations over a quasigroup. 
The design uses a quasigroup $(Q,\ast)$ of order $16$ given by the following Latin square.
\vspace{-2 mm}
\begin{small}
 \begin{eqnarray*} \label{qlw}
  \begin{array}{c|cccccccccccccccc}
\ast &0&1&2&3&4&5&6&7&8&9&a&b&c&d&e&f\\
\hline
0&5&c&1&0&2&e&9&8&f&d&3&b&7&a&4&6\\
1&f&4&3&a&8&d&6&2&5&e&1&7&b&0&c&9\\
2&6&7&d&2&0&3&f&a&9&1&e&4&c&8&b&5\\
3&8&d&7&9&f&4&0&5&2&c&b&3&1&6&e&a\\
4&4&f&0&1&d&8&7&e&c&2&a&6&9&3&5&b\\
5&9&b&e&8&a&1&5&0&6&3&d&c&4&2&7&f\\
6&a&1&c&f&9&b&2&6&0&7&4&e&d&5&3&8\\
7&e&2&9&7&c&5&1&4&d&f&6&a&0&b&8&3\\
8&7&6&8&e&3&0&4&1&b&a&2&f&5&d&9&c\\
9&2&e&b&6&5&c&a&f&8&4&7&1&3&9&d&0\\
a&b&9&2&d&1&a&c&3&7&0&8&5&f&e&6&4\\
b&0&3&4&5&6&7&8&9&a&b&c&d&e&f&1&2\\
c&3&0&f&c&7&6&d&b&1&9&5&8&2&4&a&e\\
d&1&a&5&4&b&9&e&7&3&6&f&2&8&c&0&d\\
e&d&8&6&b&4&f&3&c&e&5&9&0&a&7&2&1\\
f&c&5&a&3&e&2&b&d&4&8&0&9&6&1&f&7
\end{array}\vspace{-2 mm}
\end{eqnarray*}
\end{small}
The security of quasigroup based cryptographic algorithms also depends on the choice of the quasigroup.
An algorithm to construct cryptographically suitable quasigroups of any order $p^n$ has been given by Artamonov et. al. The above mentioned quasigroup has been generated using the algorithm given \cite{ASST}. It is polynomially complete and does not have any proper subquasigroups. The algebraic degree of vector Boolean representation of the quasigroup is $6$.  
\par
The round functions of the design consist of key mixing, an elementary string transformation over $(Q,\ast)$ for providing the confusion and again an elementary transformation over a linear quasigroup $(\mathbb{F}_2,\oplus)$  which provides diffusion. The 64-bit long round keys are generated from a master key of 128-bit length using a key scheduling algorithm, given in the subsection \ref{kgenalgo}. The keys are mixed by xoring.
The elementary  transformation over $(Q,\ast)$ is nonlinear. It is applied from the left side of $64$-bit string  in odd rounds and from right side of the string in even rounds. We denote them by \texttt{eLeft} and \texttt{eRight} for the odd and even rounds, respectively. The pictorial representation of \texttt{eLeft} and \texttt{eRight} are given in Figures \ref{figeleft} and \ref{figeright}.
\begin{figure}[!htb]
\begin{center}
\begin{tikzpicture}[node distance=0.1cm]

\node (ll) {};
\node[right = 0.4 of ll]  (a0) {$m_0$};
\node[right = 0.15 of a0] (a1) {$m_1$};
\node[right = 0.15 of a1] (a2) {$m_2$};
\node[right = 0.15 of a2] (a22) {};
\node[right = 0.70 of a22] (1a) {$\cdots$}; 
\node[right = 0.18 of 1a] (2a) {$\cdots$};
\node[right = 0.18 of 2a] (3a) {$\cdots$};
\node[right = 0.70 of 3a] (a33) {};
\node[right = 0.15 of a33] (a13) {$m_{13}$};
\node[right = 0.15 of a13] (a14) {$m_{14}$};
\node[right = 0.15 of a14] (a15) {$m_{15}$};
\node[block, fit=(a0)(a1)(a2)(a22)(1a)(2a)(3a)(a33)(a13)(a14)(a15)](aa){};

\node[below = 1.5 of a0]  (b0) {$c_0$};
\node[below = 1.5 of a1] (b1) {$c_1$};
\node[below = 1.5 of a2] (b2) {$c_2$};
\node[below = 1.5 of a22] (b22) {};
\node[below = 1.55 of 1a] (1b) {$\cdots$};
\node[below = 1.55 of 2a] (2b) {$\cdots$};
\node[below = 1.55 of 3a] (3b) {$\cdots$};
\node[below = 1.5 of a33] (b33) {};
\node[below = 1.5 of a13] (b13) {$c_{13}$};
\node[below = 1.5 of a14] (b14) {$c_{14}$};
\node[below = 1.5 of a15] (b15) {$c_{15}$};
\node[block, fit=(b0)(b1)(b2)(b22)(1b)(2b)(3b)(b33)(b13)(b14)(b15)](bb){};
\node[below = 1.6 of ll](l){$l$};

\node[ above = 0.4 of $(l.north)!0.5!(b0.north)$] (o1) {$\ast$};
\node[ above = 0.4 of $(b0.north)!0.5!(b1.north)$] (o2) {$\ast$};
\node[ above = 0.4 of $(b1.north)!0.5!(b2.north)$] (o3) {$\ast$};
\node[ above = 0.4 of $(b13.north)!0.5!(b14.north)$] (o4) {$\ast$};
\node[ above = 0.4 of $(b14.north)!0.5!(b15.north)$] (o5) {$\ast$};

\draw (l) -- (o1);
\draw (b0) -- (o2);
\draw (b1) -- (o3);
\draw (b13) -- (o4);
\draw (b14) -- (o5);
\draw[line] (o1) -- (a0);
\draw[line] (o2) -- (a1);
\draw[line] (o3) -- (a2);
\draw[line] (o4) -- (a14);
\draw[line] (o5) -- (a15);
\draw[line] (a0) -- (b0);
\draw[line] (a1) -- (b1);
\draw[line] (a2) -- (b2);
\draw[line] (a13) -- (b13);
\draw[line] (a14) -- (b14);
\draw[line] (a15) -- (b15);
\end{tikzpicture}
\caption{\texttt{eLeft}$_{(\mathbf{Q},\ast), {l}}$}
\label{figeleft}
\end{center}
\end{figure}
\begin{figure}[!htb]
\begin{center}
\begin{tikzpicture}[node distance=0.1cm]

\node (ll) {};
\node[right = 0.4 of ll]  (a0) {$m_0$};
\node[right = 0.15 of a0] (a1) {$m_1$};
\node[right = 0.15 of a1] (a2) {$m_2$};
\node[right = 0.15 of a2] (a22) {};
\node[right = 0.70 of a22] (1a) {$\cdots$}; 
\node[right = 0.18 of 1a] (2a) {$\cdots$};
\node[right = 0.18 of 2a] (3a) {$\cdots$};
\node[right = 0.70 of 3a] (a33) {};
\node[right = 0.15 of a33] (a13) {$m_{13}$};
\node[right = 0.15 of a13] (a14) {$m_{14}$};
\node[right = 0.15 of a14] (a15) {$m_{15}$};
\node[block, fit=(a0)(a1)(a2)(a22)(1a)(2a)(3a)(a33)(a13)(a14)(a15)](aa){};
\node[right = 0.3 of a15](l2){};
\node[below = 1.5 of a0]  (b0) {$c_0$};
\node[below = 1.5 of a1] (b1) {$c_1$};
\node[below = 1.5 of a2] (b2) {$c_2$};
\node[below = 1.5 of a22] (b22) {};
\node[below = 1.55 of 1a] (1b) {$\cdots$};
\node[below = 1.55 of 2a] (2b) {$\cdots$};
\node[below = 1.55 of 3a] (3b) {$\cdots$};
\node[below = 1.5 of a33] (b33) {};
\node[below = 1.5 of a13] (b13) {$c_{13}$};
\node[below = 1.5 of a14] (b14) {$c_{14}$};
\node[below = 1.5 of a15] (b15) {$c_{15}$};
\node[block, fit=(b0)(b1)(b2)(b22)(1b)(2b)(3b)(b33)(b13)(b14)(b15)](bb){};
\node[below = 1.6 of l2](l){$l'$};

\node[ above = 0.4 of $(l.north)!0.5!(b15.north)$] (o1) {$\ast$};
\node[ above = 0.4 of $(b0.north)!0.5!(b1.north)$] (o2) {$\ast$};
\node[ above = 0.4 of $(b1.north)!0.5!(b2.north)$] (o3) {$\ast$};
\node[ above = 0.4 of $(b13.north)!0.5!(b14.north)$] (o4) {$\ast$};
\node[ above = 0.4 of $(b14.north)!0.5!(b15.north)$] (o5) {$\ast$};

\draw (l) -- (o1);
\draw (b1) -- (o2);
\draw (b2) -- (o3);
\draw (b14) -- (o4);
\draw (b15) -- (o5);
\draw[line] (o1) -- (a15);
\draw[line] (o2) -- (a0);
\draw[line] (o3) -- (a1);
\draw[line] (o4) -- (a13);
\draw[line] (o5) -- (a14);
\draw[line] (a0) -- (b0);
\draw[line] (a1) -- (b1);
\draw[line] (a2) -- (b2);
\draw[line] (a13) -- (b13);
\draw[line] (a14) -- (b14);
\draw[line] (a15) -- (b15);
\end{tikzpicture}
\caption{\texttt{eRight}$_{(\mathbf{Q},\ast), {l'}}$}
\label{figeright}
\end{center}
\end{figure}
The elements $l,l'\in Q$ are commonly known as leaders.   
Since $\eL$ and $\eR$ operate over $(Q,\ast)$, the $64$-bit input divided into $16$, $4$-bit elements of $Q$. The output of the transformation is being diffused using $\texttt{eLeft}_{(\mathbb{F}_2,\oplus),1}$ and $\texttt{eRight}_{(\mathbb{F}_2,\oplus),0}$ transformations in even and odd rounds, respectively.
Note that the   $\texttt{eLeft}_{(\mathbb{F}_2,\oplus),1}$ and $\texttt{eRight}_{(\mathbb{F}_2,\oplus),0}$ are linear functions, see Figure \ref{exorleft} and \ref{exorright}. 
\vspace{-4mm}
\begin{scriptsize}
\begin{center}
\begin{figure}[!htb]
\begin{tikzpicture}[node distance=0.1cm]
\node (ll) {};

\node[right = 0.5 of ll]  (b0) {$m_0^0$};
\node[right = 0.10 of b0] (b1) {$m_0^1$};
\node[right = 0.10 of b1] (b2) {$m_0^2$};
\node[right = 0.10 of b2] (b22) {$m_0^3$};
\node[right = .70 of b22] (1b) {$\cdots$};
\node[right = 0.18 of 1b] (2b) {$\cdots$};
\node[right = 0.18 of 2b] (3b) {$\cdots$};
\node[right = 0.70 of 3b] (b33) {$m_{15}^0$};
\node[right = 0.10 of b33] (b13) {$m_{15}^1$};
\node[right = 0.10 of b13] (b14) {$m_{15}^2$};
\node[right = 0.10 of b14] (b15) {$m_{15}^3$};
\node[block, fit=(b0)(b1)(b2)(b22)(1b)(2b)(3b)(b33)(b13)(b14)(b15)](bb){};

\node[below = 1.5 of b0]  (d0) {$c_0^0$};
\node[below = 1.5 of b1] (d1) {$c_0^1$};
\node[below = 1.5 of b2] (d2) {$c_0^2$};
\node[below = 1.5 of b22] (d22) {$c_0^3$};
\node[below = 1.68 of 1b] (1d) {$\cdots$};
\node[below = 1.68 of 2b] (2d) {$\cdots$};
\node[below = 1.68 of 3b] (3d) {$\cdots$};
\node[below = 1.5 of b33] (d33) {$c_{15}^0$};
\node[below = 1.5 of b13] (d13) {$c_{15}^1$};
\node[below = 1.5 of b14] (d14) {$c_{15}^2$};
\node[below = 1.5 of b15] (d15) {$c_{15}^3$};
\node[block, fit=(d0)(d1)(d2)(d22)(1d)(2d)(3d)(d33)(d13)(d14)(d15)](dd){};
leader exor 
\node[below = 1.8 of ll](l3){$b=1$};


%
%
\node[ above = 0.4 of $(d0.north)!0.5!(d1.north)$] (eo1) {$\oplus$};
\node[ above = 0.4 of $(d1.north)!0.5!(d2.north)$] (eo2) {$\oplus$};
\node[ above = 0.4 of $(d2.north)!0.5!(d22.north)$] (eo23) {$\oplus$};
\node[ above = 0.4 of $(d33.north)!0.5!(d13.north)$] (eo24) {$\oplus$};
\node[ above = 0.4 of $(d13.north)!0.5!(d14.north)$] (eo3) {$\oplus$};
\node[ above = 0.4 of $(d14.north)!0.5!(d15.north)$] (eo4) {$\oplus$};
\node[ above = 0.44 of $(d0.north)!0.5!(l3.north)$] (eo5) {$\oplus$};
%
\draw (l3) -- (eo5);
\draw (d14) -- (eo4);
\draw (d13) -- (eo3);
\draw (d33) -- (eo24);
\draw (d2) -- (eo23);
\draw (d1) -- (eo2);
\draw (d0) -- (eo1);
\draw[line] (eo5) -- (b0);
\draw[line] (eo4) -- (b15);
\draw[line] (eo3) -- (b14);
\draw[line] (eo24) -- (b13);
\draw[line] (eo23) -- (b22);
\draw[line] (eo2) -- (b2);
\draw[line] (eo1) -- (b1);
\draw[line] (b0) -- (d0);
\draw[line] (b1) -- (d1);
\draw[line] (b2) -- (d2);
\draw[line] (b22) -- (d22);
\draw [line] (b33)-- (d33);
\draw[line] (b13) -- (d13);
\draw[line] (b14) -- (d14);
\draw[line] (b15) -- (d15);
\end{tikzpicture}
\caption{\texttt{eLeft}$_{(\mathbb{F}_2,\oplus), 1}$}\label{exorleft}
\end{figure}
\end{center}
\end{scriptsize}
\vspace{-1cm}
\begin{scriptsize}
\begin{center}
\begin{figure}[!htb]
\begin{tikzpicture}[node distance=0.1cm]
\node (ll) {};

\node[right = 0.4 of ll]  (b0) {$m_0^0$};
\node[right = 0.10 of b0] (b1) {$m_0^1$};
\node[right = 0.10 of b1] (b2) {$m_0^2$};
\node[right = 0.10 of b2] (b22) {$m_0^3$};
\node[right = .70 of b22] (1b) {$\cdots$};
\node[right = 0.18 of 1b] (2b) {$\cdots$};
\node[right = 0.18 of 2b] (3b) {$\cdots$};
\node[right = 0.70 of 3b] (b33) {$m_{15}^0$};
\node[right = 0.10 of b33] (b13) {$m_{15}^1$};
\node[right = 0.10 of b13] (b14) {$m_{15}^2$};
\node[right = 0.10 of b14] (b15) {$m_{15}^3$};
\node[block, fit=(b0)(b1)(b2)(b22)(1b)(2b)(3b)(b33)(b13)(b14)(b15)](bb){};
\node[right = 0.5 of b15](l2){};

\node[below = 1.5 of b0]  (d0) {$c_0^0$};
\node[below = 1.5 of b1] (d1) {$c_0^1$};
\node[below = 1.5 of b2] (d2) {$c_0^2$};
\node[below = 1.5 of b22] (d22) {$c_0^3$};
\node[below = 1.68 of 1b] (1d) {$\cdots$};
\node[below = 1.68 of 2b] (2d) {$\cdots$};
\node[below = 1.68 of 3b] (3d) {$\cdots$};
\node[below = 1.5 of b33] (d33) {$c_{15}^0$};
\node[below = 1.5 of b13] (d13) {$c_{15}^1$};
\node[below = 1.5 of b14] (d14) {$c_{15}^2$};
\node[below = 1.5 of b15] (d15) {$c_{15}^3$};
\node[block, fit=(d0)(d1)(d2)(d22)(1d)(2d)(3d)(d33)(d13)(d14)(d15)](dd){};
leader exor 
\node[below = 1.8 of l2](l3){$b=0$};


%
%
\node[ above = 0.4 of $(d0.north)!0.5!(d1.north)$] (eo1) {$\oplus$};
\node[ above = 0.4 of $(d1.north)!0.5!(d2.north)$] (eo2) {$\oplus$};
\node[ above = 0.4 of $(d2.north)!0.5!(d22.north)$] (eo23) {$\oplus$};
\node[ above = 0.4 of $(d33.north)!0.5!(d13.north)$] (eo24) {$\oplus$};
\node[ above = 0.4 of $(d13.north)!0.5!(d14.north)$] (eo3) {$\oplus$};
\node[ above = 0.4 of $(d14.north)!0.5!(d15.north)$] (eo4) {$\oplus$};
\node[ above = 0.44 of $(d15.north)!0.5!(l3.north)$] (eo5) {$\oplus$};
%
\draw (l3) -- (eo5);
\draw (d15) -- (eo4);
\draw (d14) -- (eo3);
\draw (d13) -- (eo24);
\draw (d22) -- (eo23);
\draw (d2) -- (eo2);
\draw (d1) -- (eo1);
\draw[line] (eo5) -- (b15);
\draw[line] (eo4) -- (b14);
\draw[line] (eo3) -- (b13);
\draw[line] (eo24) -- (b33);
\draw[line] (eo23) -- (b2);
\draw[line] (eo2) -- (b1);
\draw[line] (eo1) -- (b0);
\draw[line] (b0) -- (d0);
\draw[line] (b1) -- (d1);
\draw[line] (b2) -- (d2);
\draw[line] (b22) -- (d22);
\draw [line] (b33)-- (d33);
\draw[line] (b13) -- (d13);
\draw[line] (b14) -- (d14);
\draw[line] (b15) -- (d15);
\end{tikzpicture}
\caption{\texttt{eRight}$_{(\mathbb{F}_2,\oplus), 0}$}\label{exorright}
\end{figure}
\end{center}
\end{scriptsize}
The mathematical representation of round functions:
\begin{itemize}
\item xoring of round keys:
\begin{eqnarray*}
\kM :  \mathbb{F}_2^{64}\times \mathbb{F}_2^{64} &\rightarrow & \mathbb{F}_2^{64},\\
      ((k_0,\ldots,k_{63},~a_0,\ldots,a_{63})&\mapsto&(k_0\oplus a_0,\ldots,k_{63}\oplus a_{63})
\end{eqnarray*} 
\item confusion transformations:
\begin{eqnarray*}
\eL_{(Q,\ast),l}:  Q^{16}\rightarrow Q^{16},\quad a_0,\ldots,a_{15}\mapsto b_0,\ldots,b_{15}, \\
 \quad b_0=l\ast a_0,\quad b_{i}=b_{i-1}\ast a_{i},\quad i=1,\ldots, 15\\[2mm]
%
\eR_{(Q,\ast),l'}: Q^{16}\rightarrow Q^{16},\quad a_0,\ldots,a_{15}\mapsto b_0,\ldots,b_{15}, \\
 \quad b_{15}=l'\ast a_{15},\quad b_{i}=b_{i+1}\ast a_{i},\quad i=14,\ldots,0
\end{eqnarray*}
\item diffusion transformations:
\begin{eqnarray*}
\eL_{(\mathbb{F}_2,\oplus),1}:  \mathbb{F}_2^{64}\rightarrow \mathbb{F}_2^{64},\quad
\eR_{(\mathbb{F}_2,\oplus),0}: \mathbb{F}_2^{64}\rightarrow \mathbb{F}_2^{64}
\end{eqnarray*}
\end{itemize}
Except $16^{th}$ round, the odd and even round functions are 
\vspace{-0.2cm}
\begin{eqnarray*}
\text{\RO} &=&\texttt{eRight}_{(\mathbb{F}_2,\oplus),0}\circ \eL_{(Q,\ast),l}\circ \kM \\
\text{\RE} &=&\texttt{eLeft}_{(\mathbb{F}_2,\oplus),1}\circ \eR_{(Q,\ast),l}\circ \kM
\end{eqnarray*}
The last round does not contain diffusion transformation. The final output is xored with $17^{th}$ 64-bit key. 
The leader $l$ used in $\eL$ and $\eR$ transformations are first and last 4-bit of the odd and even round keys, respectively. 
The pseudo-code of the  encryption and decryption functions are given in Algorithm \ref{enc} and \ref{dec}.
\begin{algorithm}[H]
\caption{Encryption Algorithm}
\label{enc}
\begin{algorithmic}[1]
\REQUIRE Plaintext $m$ and $64$-bit round keys $rk_0,\ldots,rk_{16}$
\ENSURE  Ciphertext $c$
\STATE $c:=m$
\FOR{$1\leq i \leq 16$}
\STATE $c=\kM(rk_{i-1}, c)$
\IF{$i$ is odd}
\STATE $c=\eL_{{(Q,\ast),l}}(c)$
\STATE $c=\eR_{(\mathbb{F}_2,\oplus),0}(c)$
\ELSE
\STATE $c=\eR_{{(Q,\ast),l}}(c)$
\IF{$i\neq 16$}
\STATE $c=\eL_{(\mathbb{F}_2,\oplus),1}(c)$
\ENDIF
\ENDIF
\ENDFOR
\STATE $c=\kM(rk_{16}, c)$
\RETURN $c$
\end{algorithmic}
\end{algorithm}
\begin{algorithm}[H]
\caption{Decryption Algorithm}
\label{dec}
\begin{algorithmic}[1]
\REQUIRE Ciphertext $c$ and $64$-bit round keys $rk_{16},\ldots,rk_{0}$
\ENSURE  Plaintext $m$
\STATE $m:=c$
\STATE $m=\texttt{kXor}(rk_{16},m)$
\FOR{$1\leq i \leq 16$}
\IF{$i$ is odd}
\IF{$i\neq 1$}
\STATE $m=\dL_{(\mathbb{F}_2,\backslash_{\oplus),1}}(m)$
\ENDIF
\STATE $m=\dR_{l,(Q,\backslash_{\ast)}}(m)$
\ELSE
\STATE $m=\dR_{(\mathbb{F}_2,\backslash_{\oplus),0}}(m)$
\STATE $m=\dL_{l,(Q,\backslash_{\ast)}}(m)$
\ENDIF
\STATE $m=\kM(rk_{16-i},m)$
\ENDFOR
\RETURN $m$
\end{algorithmic}
\end{algorithm}
In the decryption process, the round function operates in the reverse order. We use different string transformation over (inverse) conjugate of $(Q,\ast)$, denoted by $(Q,\backslash_\ast)$. We denote by $\dL$ and $\dR$, the string transformations applied from the left and right sides of the string. They are the inverse transformations of  $\eL$ and $\eR$, respectively. 
The pictorial representations of $\dL$ and $\dR$ are given in Figure \ref{dleft} and \ref{dright}.

\begin{figure}[!htb]
\begin{center}

\begin{tikzpicture}[node distance=0.1cm]

\node (ll) {$l$};
\node[right = 0.3 of ll](s0){$\backslash _{\ast}$};
\node[right = 0.3 of s0]  (a0) {$c_0$};
\node[right = 0.3 of a0] (s1) {$\backslash _{\ast}$};
\node[right = 0.3 of s1] (a1) {$c_1$};
\node[right = 0.2 of a1] (a22) {};
\node[right = 0.20 of a22] (1a) {$\cdots$}; 
\node[right = 0.15 of 1a] (2a) {$\cdots$};
\node[right = 0.15 of 2a] (3a) {$\cdots$};
\node[right = 0.20 of 3a] (a33) {};
\node[right = 0.3 of a33] (a14) {$c_{14}$};
\node[right = 0.3 of a14] (s15) {$\backslash _{\ast}$};
\node[right = 0.3 of s15] (a15) {$c_{15}$};
\node[block, fit=(a0)(a1)(a22)(1a)(2a)(3a)(a33)(a14)(a15)](aa){};

\node[below = 1 of a0]  (b0) {$m_0$};
\node[below = 1 of a1] (b1) {$m_1$};
\node[below = 1 of a22] (b22) {};
\node[below = 1.05 of 1a] (1b) {$\cdots$};
\node[below = 1.05 of 2a] (2b) {$\cdots$};
\node[below = 1.05 of 3a] (3b) {$\cdots$};
\node[below = 1 of a33] (b33) {};
\node[below = 1 of a14] (b14) {$m_{14}$};
\node[below = 1 of a15] (b15) {$m_{15}$};
\node[block, fit=(b0)(b1)(b22)(1b)(2b)(3b)(b33)(b14)(b15)](bb){};


\draw[line] (s0) -- (a0);
\draw[line] (s1) -- (a1);
\draw[line] (s15) -- (a15);
\draw(ll)--(s0);
\draw(a0)--(s1);
\draw(a14)--(s15);
\draw[line] (a0) -- (b0);
\draw[line] (a1) -- (b1);
\draw[line] (a14) -- (b14);
\draw[line] (a15) -- (b15);
\end{tikzpicture}
\caption{\texttt{dLeft}$_{(\mathbb{Q},\backslash_\ast),l}$}\label{dleft}
\end{center}
\end{figure}

%
\begin{figure}[!htb]
\begin{center}

\begin{tikzpicture}[node distance=0.1cm]

\node (ll) {};
\node[right = 0.3 of ll]  (a0) {$c_0$};
\node[right = 0.3 of a0] (s1) {$\backslash _{\ast}$};
\node[right = 0.3 of s1] (a1) {$c_1$};
\node[right = 0.2 of a1] (a22) {};
\node[right = 0.20 of a22] (1a) {$\cdots$}; 
\node[right = 0.15 of 1a] (2a) {$\cdots$};
\node[right = 0.15 of 2a] (3a) {$\cdots$};
\node[right = 0.20 of 3a] (a33) {};
\node[right = 0.3 of a33] (a14) {$c_{14}$};
\node[right = 0.3 of a14] (s15) {$\backslash _{\ast}$};
\node[right = 0.3 of s15] (a15) {$c_{15}$};
\node[right = 0.3 of a15] (s0) {$\backslash _{\ast}$};
\node[right = 0.3 of s0] (l) {$l'$};
\node[block, fit=(a0)(a1)(a22)(1a)(2a)(3a)(a33)(a14)(a15)](aa){};

\node[below = 1 of a0]  (b0) {$m_0$};
\node[below = 1 of a1] (b1) {$m_1$};
\node[below = 1 of a22] (b22) {};
\node[below = 1.05 of 1a] (1b) {$\cdots$};
\node[below = 1.05 of 2a] (2b) {$\cdots$};
\node[below = 1.05 of 3a] (3b) {$\cdots$};
\node[below = 1 of a33] (b33) {};
\node[below = 1 of a14] (b14) {$m_{14}$};
\node[below = 1 of a15] (b15) {$m_{15}$};
\node[block, fit=(b0)(b1)(b22)(1b)(2b)(3b)(b33)(b14)(b15)](bb){};




\draw[line] (s0) -- (a15);
\draw[line] (s1) -- (a0);
\draw[line] (s15) -- (a14);
\draw(l)--(s0);
\draw(a1)--(s1);
\draw(a15)--(s15);

\draw[line] (a0) -- (b0);
\draw[line] (a1) -- (b1);
\draw[line] (a14) -- (b14);
\draw[line] (a15) -- (b15);
\end{tikzpicture}
\caption{\texttt{dRight}$_{(\mathbb{Q},\backslash_\ast), l'}$}\label{dright}
\end{center}
\end{figure}
In the first round, ciphertext is xored with $rk_{16}$,  and apply  $\dR_{l,(Q,\backslash_{\ast})}$ transformation and again xored with $rk_{15}$. After the first round, the round functions
\begin{eqnarray*}
\text{\RO}&=& \texttt{kXor} \circ \dR_{(Q,\backslash_{\ast}),l}\circ \texttt{dLeft}_{(\mathbb{F}_2,\backslash_{\oplus}),1}, \\
\text{\RE}&=& \texttt{kXor} \circ \dL_{(Q,\backslash_{\ast}),l}\circ \texttt{dRight}_{(\mathbb{F}_2,\backslash_{\oplus}),0}
\end{eqnarray*}
The schematic diagram of the encryption algorithm is given in Figure \ref{complete}.
\begin{center}
\begin{figure}[h!]
\begin{tikzpicture}[node distance=0.1cm]
\node  (b0) {$m_0$};
\node[right = 0.10 of b0] (b1) {$m_1$};
\node[right = 0.10 of b1] (b2) {$m_2$};
\node[right = 0.10 of b2] (b22) {$m_3$};
\node[right = 0.1 of b22] (1b) {$\cdot$};
\node[right = 0.05 of 1b] (2b) {$\cdot$};
\node[right = 0.05 of 2b] (3b) {$\cdot$};
\node[right = 0.1 of 3b] (b33) {$m_{60}$};
\node[right = 0.10 of b33] (b13) {$m_{61}$};
\node[right = 0.10 of b13] (b14) {$m_{62}$};
\node[right = 0.10 of b14] (b15) {$m_{63}$};
\node[ fit=(b0)(b1)(b2)(b22)(1b)(2b)(3b)(b33)(b13)(b14)(b15)](bb){};
\node[below = 0.2 of b0]  (d0) {};
\node[below = 0.2 of b1] (d1) {};
\node[below = 0.2 of b2] (d2) {};
\node[below = 0.2 of b22] (d22) {};
\node[below = 0.2 of 2b] (2d) {};
\node[below = 0.2 of b33] (d33) {};
\node[below = 0.2 of b13] (d13) {};
\node[below = 0.2 of b14] (d14) {};
\node[below = 0.2 of b15] (d15) {};
\node[ fit=(d0)(d1)(d2)(d22)(d33)(d13)(d14)(d15)](dd){};

\node[rec, below = 0.1 of bb](ke0){xoring of subkey $rk_0$};


\node[below = 0.2 of 2d] (2e) {};

\node[below = 0.47 of 2e] (2f) {};
\draw[line] (2e)--(2f);

\node[rec, below = 0.6 of ke0](el1){eLeft$_{(Q,\ast),l}$  };
\node[below = 0.21 of 2f] (2ee) {};

\node[below = 0.5 of 2ee] (2ff) {};
\draw[line] (2ee)--(2ff);

\node[rec, below = 0.57 of el1](er1){ eRight$_{(\mathbb{F}_2,\oplus),0}$ };

\node[below = 0.22 of 2ff] (2g) {};

\node[below = 0.01 of 2b](ar0){};
\node[below = 0.01 of 2g] (ar1) {};

\node[below = 0.69 of 2g]  (2q) {};
\draw[line] (2g) -- (2q);

\node[rec, below = 0.7 of er1](ke1){xoring of subkey $rk_1$};

\node[below = 0.1 of 2q] (2r) {};
\node[below = 0.51 of 2r] (2fff) {};
\draw[line] (2r)--(2fff);
\node[above = 0.01 of 2q] (ar2) {};
\node[rec, below = 0.55 of ke1](er2){eRight$_{(Q,\ast),l^{'}}$};

\node[below = 1.37 of 2q] (2i) {};
\node[below = 1.4 of 2q] (xx) {};
\node[below = 0.5 of xx] (xxx) {};
\draw[line] (xx) -- (xxx);
\node[rec, below = 0.55 of er2](el2){eLeft$_{(\mathbb{F}_2,\oplus),1}$  };

\node[below = 7 of d2] (vl1) {$\vdots$};
\node[below = 1.35 of 2i] (vv3) {$\vdots$};
\node[below = 7 of d13] (vr1) {$\vdots$};

\node[below = 0.1 of vv3] (2qp) {};
%
\node[rec, below = 1.1 of el2](ke15){xoring of subkey $rk_{15}$};
%
\node[below = 0.26 of 2qp] (2rr) {};
\node[above = 1.1 of 2rr] (ar3) {};
\node[below = 2 of 2i] (ar4) {};

\node[rec, below = 0.55 of ke15](er3){eRight$_{(Q,\ast),l^{''}}$ };
\node[below = 0.5 of 2rr] (2ff) {};
\draw[line](2rr)--(2ff);

\node[below = 0.25 of 2ff] (2ii) {};
\node[below = 0.67 of 2ii] (2n) {};

\node[below = 0.05 of 2ii] (ar5) {};

\draw[line](2ii)--(2n);
\node[rec, below = 0.74 of er3](ke16){xoring of subkey $rk_{16}$};
%
\node[below = 11.1 of d0]  (p0) {};
\node[below = 11.1 of d1] (p1) {};
\node[below = 11.1 of d2] (p2) {};
\node[below = 11.1 of d22] (p22) {};
\node[below = 11.1 of d33] (p33) {};
\node[below = 11.1 of d13] (p13) {};
\node[below = 11.1 of d14] (p14) {};
\node[below = 11.1 of d15] (p15) {};

\node[below = 0.2 of p0]  (s0) {$c_0$};
\node[below = 0.2 of p1] (s1) {$c_1$};
\node[below = 0.2 of p2] (s2) {$c_2$};
\node[below = 0.2 of p22] (s22) {$c_3$};
\node[below = 12.15 of 1b] (1s) {$\cdot$};
\node[below = 12.15 of 2b] (2s) {$\cdot$};
\node[below = 12.15 of 3b] (3s) {$\cdot$};
\node[below = 0.2 of p33] (s33) {$c_{60}$};
\node[below = 0.2 of p13] (s13) {$c_{61}$};
\node[below = 0.2 of p14] (s14) {$c_{62}$};
\node[below = 0.2 of p15] (s15) {$c_{63}$};

%
\draw [decorate,
    decoration = {calligraphic brace,mirror,amplitude= 8pt,raise=128pt},line width=0.8] (ar0) --  (ar1)
   node[pos=0.5,left=133pt,black]{ Round $1$};
   
 \draw [decorate,     
    decoration = {calligraphic brace,amplitude= 8pt,raise=145pt},,line width=0.8] (ar2) --  (ar3)
    node[pos=0.5,right=150pt,black]{ Round $2$};
    
     \draw [decorate,     
    decoration = {calligraphic brace,amplitude= 8pt,raise=145pt},,line width=0.8] (ar4) --  (ar5)
    node[pos=0.5,right=150pt,black]{ Round $16$};
\end{tikzpicture}
\caption{\texttt{Encryption Algorithm of the Design}}\label{complete}
\end{figure}
\end{center}

\begin{remark}
As, the conjugates of a polynomially complete finite quasigroup is also polynomial complete. So, the quasigroup used in the decryption $(Q,\backslash_{\ast})$ remains polynomially complete which ensures the nonlinearity of the decryption function \cite{normal}.  
\end{remark}
Now, we present our round keys generation algorithm from a 128-bit master key.
\subsection{Key Scheduling Algorithm}\label{kgenalgo}
Each round of the design requires 64-bit long keys. We generate $17$,  64-bit round keys in the following steps.  \\\noindent
First, a master $\texttt{key}=k_0,k_1,\ldots,k_{31}$ is mixed with a $64$-bit $\texttt{iv}=v_0,v_1,\ldots,v_{15}$ and   a $64$-bit fixed string $``15,14,\ldots,0"$ using  Algorithm \ref{kMix}. 
\begin{algorithm}[H]
\caption{Key Mixing}
\label{kMix}
\begin{algorithmic}[1]
\REQUIRE $\mathsf{key}=(k_0,k_1,\ldots,k_{31}),~\mathsf{iv}=v_0,v_1,\ldots ,v_{15}$
\ENSURE  Mixed key with $\mathsf{iv}$
\STATE   $\mathsf{s}=k_0,k_1,\ldots ,k_{31},v_0,v_1,\ldots ,v_{15},15,14,\ldots,0$
\STATE   $\mathsf{a}=\mathsf{s}$
\FOR{$1\leq i \leq 64$}
\IF{$i$ is odd}
\STATE $\mathsf{a}=\eL_{(Q,\ast),\mathsf{s}_{64-i}}(\mathsf{a})$
\ELSE
\STATE $\mathsf{a}=\eR_{(Q,\ast),\mathsf{s}_{64-i}}(\mathsf{a})$
\ENDIF
\ENDFOR
\RETURN $\mathsf{a}$
\end{algorithmic}
\end{algorithm}
Then, the round keys $rk_0,rk_1,\ldots,rk_{16}$ are generated from the output of the previous algorithm in the following way.  
\begin{algorithm}[H]
\caption{Rounds Key Generation}
\label{kgs}
\begin{algorithmic}[1]
\REQUIRE Mixed key with $\mathsf{iv}$: $\mathsf{a} = \mathsf{a}_0,\ldots, \mathsf{a}_{63}$ (Output of Key Mixing)
\ENSURE Round keys $rk$
\STATE $\mathsf{s}=\underbrace{0,1,2,\ldots ,15}_{1},\ldots,\underbrace{0,1,2,\ldots ,15}_{34}$
\STATE $\l=\mathsf{s}$

\FOR{$1\leq i \leq 64$}
\IF{$i$ is odd}
\STATE $\l=\eL_{(Q,\ast),\mathsf{a}_{i-1}}(l)$
\ELSE
\STATE $\l=\eR_{(Q,\ast),\mathsf{a}_{i-1}}(l)$
\ENDIF
\ENDFOR
\FOR{$0\leq i\leq 16$}
\STATE $rk_{i}=(\l_{32*i+0},\l_{32*i+2},\ldots,\l_{32*i+30})$
\ENDFOR
\RETURN $rk=(rk_0,\ldots,rk_{16})$
\end{algorithmic}
\end{algorithm}
\section{Security Analysis of the Design}\label{s3}
\subsection{Linear Cryptanalysis}
The quasigroup binary operation $\ast$ is nonlinear. The string transformations $\eL$ and $\eR$ can be considered as a confusion layer of $16$ Sboxes of type $4\times 4$.  There could be $16$ different Sboxes for 16 different leaders from $Q$. The binary operation $\ast$ for a fixed leader $l\in Q$ provides a Sbox
$$S_{l}:Q\rightarrow Q, \quad x\mapsto l\ast x, \quad x\in Q $$ 
Clearly, the rows of the quasigroup provide Sbox as $S_0,S_1,\ldots,S_{15}$.
The (nonlinear) Sbox layer is output dependent that is the output of an Sboxes decides which Sbox $S_i$ will be considered next. 
\par
In the linear cryptanalysis  \cite{lc}, one tries to approximate the nonlinear component of the design by some linear relation between the output bits at some round with the plaintext bits and key bits. Here we show that the standard approach of getting linear approximation relation  is not applicable in the design even at the end of the second round. So, the design is resistant to linear attack. 
\par
Let $x_0,x_1,\ldots,x_{63}$ is a plaintext and $k_0^{1},\ldots,k_{63}^{1}$ is the first round key. Then, for an arbitrary nonzero $4$-bit input-output masks $(a_0,a_1,a_2,a_3)$ and $(b_0,b_1,b_2,b_3)$ corresponding to $16^{th}$ Sbox we have 
\begin{eqnarray}
\mathbf{
b_0y_{60}\oplus\cdots \oplus b_3y_{63}=a_0(x_{60}\oplus k^1_{60})\oplus \cdots\oplus a_3(x_{63}\oplus k^1_{63})
}
\end{eqnarray}  
where $y_0,\ldots,y_{63}$ represent the output of the Sbox layer. If $z_0,\ldots, z_{63}$ denote the output after diffusion layer, we get the following relations for the last 4-bit  
\begin{eqnarray}
\hspace{4 mm}\mathbf{z_{60}=y_{60}\oplus z_{61},~ z_{61}=y_{61}\oplus z_{62},~  z_{62}=y_{62}\oplus z_{63},~  z_{63}=y_{63},} 
\end{eqnarray}
Similarly, if $u_0,\ldots,u_{63}$ denote the output after second round Sbox layer, we get relations for an arbitrary input-output masks 
$(a'_0,a'_1,a'_2,a'_3)$ and $(b'_0,b'_1,b'_2,b'_3)$,
\begin{eqnarray}
\mathbf{b'_0u_{60}\oplus \cdots \oplus b'_3u_{63}=a'_{0}(z_{60}\oplus k^2_{60})\oplus \dots \oplus a'_{3}(z_{63}\oplus k^2_{63})}
\end{eqnarray}

If $v_0,\ldots,v_{63}$ denote the output of diffusion layer in second round, then
\begin{eqnarray}\label{eq5}
\mathbf{
v_{60}=\oplus _{i=0}^{60}u_i\oplus 1,~\cdots,  ~ v_{63}=\oplus _{i=0}^{63}u_i\oplus 1
}
\end{eqnarray}
Figure \ref{lc} depicts the above description.
%
\begin{figure}[!htb]
\begin{center}
\begin{scriptsize}
\begin{tikzpicture}[node distance=0.1cm]

\node (ll) {};
\node[right = 0.7 of ll]  (a0) {};
\node[right = 0.7 of a0] (am0) {};
\node[right = 0.7 of am0] (1a) {$\cdots$};
\node[right = 1.3 of 1a] (a15) {$\cdots$};


\node[right = 1.25 of am0] (1am) {$k^1_{60}\oplus x_{60} $};
\node[right = 2.45 of am0] (2amm) {};
\node[right = 3.1 of am0] (3amm) {};
\node[right = 3.1 of am0] (4am) {$k^1_{63} \oplus x_{63}$};
\node[above = 0.9 of a0] (a0ma) {$x_{0} $};
\node[above = 0.9 of am0] (am0a) {$~~~~~~~~~\cdots $};
\node[above = 0.8 of 1am] (1ama) {$x_{60} $};
\node[above = 1.19 of 2amm] (2ama) {};
\node[above = 0.9 of a15] (15ama) {$\cdots $};
\node[above = 1.19 of 3amm] (3ama) {};
\node[above = 0.8 of 4am] (4ama) {$x_{63}$};

\node[below = 0.3 of 1ama] (1amab) {};
\node[below = 0.3 of 2ama] (2amab) {};
\node[below = 0.3 of 3ama] (3amab) {};
\node[below = 0.3 of 4ama] (4amab) {};

\node[below = 1.35 of 2ama] (2am) {};
\node[below = 1.35 of 3ama] (3am) {};

\node[below = 0.37 of 1am] (1sm) {};
\node[below = 0.37 of 2am] (2sm) {};
\node[below = 0.37 of 3am] (3sm) {};
\node[below = 0.37 of 4am] (4sm) {};

\node[below = 0.09 of 1sm] (1bm) {};
\node[below = 0.09 of 2sm] (2bm) {};
\node[below = 0.09 of 3sm] (3bm) {};
\node[below = 0.09 of 4sm] (4bm) {};

\node[below = 0.31 of 1bm] (1bbm) {$y_{60}$};
\node[below = 0.3 of 2bm] (2bbm) {};
\node[below = 0.3 of 3bm] (3bbm) {};
\node[below = 0.31 of 4bm] (4bbm) {$y_{63}$};

\node[fit=(a0)(1a)(am0)(1am)(2am)(3am)(4am)](aa){};

\node[below = 1.4 of a0]  (b0) {};
\node[below = 1.35  of 1a] (1b) {$\cdots$};
\node[below = 1.4 of am0] (bm0) {};
\node[below = 1.3  of a15] (b15) {$\cdots $};
\node[ fit=(b0)(1b)(bm0)(b15)](bb){};
\node[right = 1.4 of b15](l2){};

\node[sqr1,draw, below = 0.5 of a0] (s0) {S};
\node[ below = 0.46 of 1a] (1s) {$\cdots$};
\node[sqr1,draw, below = 0.5 of am0] (s1) {S};
\node[sqrb,draw, below = 0.42 of a15] (s4) {\textbf{S}};
\node[ fit=(s0)(s1)(s4)](ss){};
\node[below = 0.8 of b0]  (d0) {};
\node[below = 0.65 of 1b] (1d) {$\cdots$};
\node[below = 0.8 of bm0] (dm0) {};
\node[below = 0.7 of b15] (d15) {$\cdots$};
\node[below = 0.65 of 1bbm] (1ddm) {$z_{60}$};
\node[below = 0.65 of 2bbm] (2ddm) {};
\node[below = 0.65 of 3bbm] (3ddm) {};
\node[below = 0.65 of 4bbm] (4ddm) {$z_{63}$};

\node[ fit=(d0)(1d)(dm0)(d15)(1ddm)(4ddm)](dd){};
\node[below = 0.75 of l2](l3){$0$};
\node[rec1, below = 1.8 of ss](key1){kXor};
\node[rec1, above = 3.5 of key1](key0){ kXor}; 
\node[ above = 0.15 of $(dm0.north)!0.5!(d0.north)$] (eo1) {$\oplus$};
\node[ below = 0.53 of $(4bbm.north)!0.5!(l2.north)$] (eo3) {$\oplus$};


\draw (l3) -- (eo3);
\draw (dm0) -- (eo1);
\draw[line] (eo3) -- (4bbm);
\draw[line] (eo1) -- (b0);
\draw[line] (b0) -- (d0);
\draw[line] (bm0) -- (dm0);
\draw[line] (1bbm) -- (1ddm);
\draw[line] (4bbm) -- (4ddm);

\draw[line] (a0) -- (s0);
\draw[line] (s0) -- (b0);
\draw[line] (am0) -- (s1);
\draw[line] (s1) -- (bm0);


\draw[line] (1am) -- (1sm);
\draw[line] (2am) -- (2sm);
\draw[line] (3am) -- (3sm);
\draw[line] (4am) -- (4sm);

\draw[line] (1bm) -- (1bbm);
\draw[line] (2bm) -- (2bbm);
\draw[line] (3bm) -- (3bbm);
\draw[line] (4bm) -- (4bbm);

\node[below = 1.2 of d0]  (a0_2) {};
\node[below = 1.2 of dm0] (am0_2) {};
\node[below = 0.9 of 1d] (1a_2) {$\cdots$};
\node[below = 0.9 of d15] (a15_2) {$\cdots$};

\node[below = 3.4 of 1am] (1am_2) {$k^2_{60} \oplus z_{60} $};
\node[below = 3.4 of 2amm] (2amm_2) {};
\node[below = 3.4 of 3amm] (3amm_2) {};
\node[below = 3.4 of 4am] (4am_2) {$k^2_{63}\oplus z_{63}$};

\node[below = 3.72 of 2am] (2am_2) {};
\node[below = 3.72 of 3am] (3am_2) {};

\node[below = 0.36 of 1am_2] (1sm_2) {};
\node[below = 0.36 of 2am_2] (2sm_2) {};
\node[below = 0.36 of 3am_2] (3sm_2) {};
\node[below = 0.36 of 4am_2] (4sm_2) {};

\node[below = 0.18 of 1sm_2] (1bm_2) {};
\node[below = 0.18 of 2sm_2] (2bm_2) {};
\node[below = 0.18 of 3sm_2] (3bm_2) {};
\node[below = 0.18 of 4sm_2] (4bm_2) {};

\node[below = 0.31 of 1bm_2] (1bbm_2) {$u_{60}$};
\node[below = 0.3 of 2bm_2] (2bbm_2) {};
\node[below = 0.3 of 3bm_2] (3bbm_2) {};
\node[below = 0.31 of 4bm_2] (4bbm_2) {$u_{63}$};

\node[fit=(a0_2)(1a_2)(am0_2)(1am_2)(2am_2)(3am_2)(4am_2)](aa_2){};

\node[below = 1.3 of a0_2]  (b0_2) {};
\node[below = 1.3  of 1a_2] (1b_2) {$\cdots$};
\node[below = 1.3  of am0_2] (bm0_2) {};
\node[below = 1.3  of a15_2] (b15_2) {$\cdots $};
\node[ fit=(b0_2)(1b_2)(bm0_2)(b15_2)](bb_2){};
\node[left = 0.7 of b0_2](l2_2){};

\node[sqr1,draw, below = 0.4 of a0_2] (s0_2) {S};
\node[ below = 0.46 of 1a_2] (1s) {$\cdots$};
\node[sqr1,draw, below = 0.4 of am0_2] (s1_2) {S};
\node[sqrb,draw, below = 0.42 of a15_2] (s4_2) {\textbf{S}};

\node[below = 0.7 of b0_2]  (d0_2) {};
\node[below = 0.75 of 1b_2] (1d_2) {$\cdots$};
\node[below = 0.7 of bm0_2] (dm0_2) {};
\node[below = 0.7 of b15_2] (d15_2) {};
\node[below = 0.65 of 1bbm_2] (1ddm_2) {$v_{60}$};
\node[below = 0.65 of 2bbm_2] (2ddm_2) {};
\node[below = 0.65 of 3bbm_2] (3ddm_2) {};
\node[below = 0.56 of 4bbm_2] (4ddm_2) {$v_{63}= \oplus _{i=0}^{63}u_i \oplus 1$};
\node[ fit=(d0_2)(1d_2)(dm0_2)(d15_2)(1ddm_2)(4ddm_2)](dd_2){};
\node[below = 0.7 of l2_2](l3_2){$1$};
\node[ above = 0.15 of $(dm0_2.north)!0.5!(d0_2.north)$] (eo1_2) {$\oplus$};
\node[ below = 0.4 of $(b0_2.north)!0.5!(l2_2.north)$] (eo3_2) {$\oplus$};

\draw (l3_2) -- (eo3_2);
\draw (d0_2) -- (eo1_2);
\draw[line] (eo3_2) -- (b0_2);
\draw[line] (eo1_2) -- (bm0_2);
\draw[line] (b0_2) -- (d0_2);
\draw[line] (bm0_2) -- (dm0_2);
\draw[line] (1bbm_2) -- (1ddm_2);
\draw[line] (4bbm_2) -- (4ddm_2);

\draw[line] (a0_2) -- (s0_2);
\draw[line] (s0_2) -- (b0_2);
\draw[line] (am0_2) -- (s1_2);
\draw[line] (s1_2) -- (bm0_2);

\draw[line] (1am_2) -- (1sm_2);
\draw[line] (2am_2) -- (2sm_2);
\draw[line] (3am_2) -- (3sm_2);
\draw[line] (4am_2) -- (4sm_2);

\draw[line] (1bm_2) -- (1bbm_2);
\draw[line] (2bm_2) -- (2bbm_2);
\draw[line] (3bm_2) -- (3bbm_2);
\draw[line] (4bm_2) -- (4bbm_2);

\end{tikzpicture}
\caption{\texttt{Linear Cryptanalysis}}\label{lc}
\end{scriptsize}
\end{center}
\vspace{-.7 cm}
\end{figure}
In the above equation, all the intermediate variables at the end of second round are involved which cannot be eliminated in the succeeding rounds in terms of plaintext bits and key variables. Therefore, the linear approximation relations cannot be obtained using the standard approach. Hence, the design is resistant to the linear attack. {Note that the above description remains same even if we consider non-linear layer as $8\times 4$ Sbox (defined in the next subsection)}. \vspace{-2cm}
\subsection{Differential Cryptanalysis}
Any block cipher design should resist the  differential cryptanalysis{ \cite{jofc-1991-14114}}. It should not be possible to make a differential distinguisher with greater   probability than $\frac{1}{2^n}$, where $n$ is the size of  a block. Instead of standard Sboxes, we use quasigroup string transformation $\eL$ and $\eR$ over $(Q,\ast)$ to provide the nonlinearity in the design. 
We can consider binary $\ast$ operation of the design as an Sbox of $8\times 4$ type 
$$S:Q\times Q\rightarrow Q, \quad (l,x) \mapsto l\ast x, \quad l,x\in Q $$
The $\eL$ and $\eR$ maps can be seen as the layers having 16 Sboxes, where $4$ in $8$ input bits of the Sbox comes from the $4$-bit output of previous Sbox. For the first Sbox in the odd and even rounds, this 4-bit are taken from first and last 4-bit of the round keys, respectively (see Figure \ref{lsbox}). The output of one Sbox goes to the input of next Sbox, the Sbox layer is ``output dependent". 
\begin{figure}[!htb]
\begin{center}
\begin{tikzpicture}[node distance=0.1cm]

\node (ll) {};
\node[right = 0.4 of ll]  (a0) {$x_0$};
\node[right = 0.15 of a0] (a1) {$x_1$};
\node[right = 0.15 of a1] (a2) {$x_2$};
\node[right = 0.15 of a2] (a22) {};
\node[right = 0.70 of a22] (1a) {$\cdots$}; 
\node[right = 0.18 of 1a] (2a) {$\cdots$};
\node[right = 0.18 of 2a] (3a) {$\cdots$};
\node[right = 0.70 of 3a] (a33) {};
\node[right = 0.15 of a33] (a13) {$x_{13}$};
\node[right = 0.15 of a13] (a14) {$x_{14}$};
\node[right = 0.15 of a14] (a15) {$x_{15}$};
\node[block, fit=(a0)(a1)(a2)(a22)(1a)(2a)(3a)(a33)(a13)(a14)(a15)](aa){};

\node[below = 1.5 of a0]  (b0) {$y_0$};
\node[below = 1.5 of a1] (b1) {$y_1$};
\node[below = 1.5 of a2] (b2) {$y_2$};
\node[below = 1.5 of a22] (b22) {};
\node[below = 1.55 of 1a] (1b) {$\cdots$};
\node[below = 1.55 of 2a] (2b) {$\cdots$};
\node[below = 1.55 of 3a] (3b) {$\cdots$};
\node[below = 1.5 of a33] (b33) {};
\node[below = 1.5 of a13] (b13) {$y_{13}$};
\node[below = 1.5 of a14] (b14) {$y_{14}$};
\node[below = 1.5 of a15] (b15) {$y_{15}$};
\node[block, fit=(b0)(b1)(b2)(b22)(1b)(2b)(3b)(b33)(b13)(b14)(b15)](bb){};
\node[below = 1.6 of ll](l){$l$};

\node[sqr,draw, below = 0.55 of a0] (s0) {S};
\node[sqr,draw, below = 0.55 of a1] (s1) {S};
\node[sqr,draw, below = 0.55 of a2] (s2) {S};
\node[sqr,draw, below = 0.55 of a13] (s13) {S};
\node[sqr,draw, below = 0.55 of a14] (s14) {S};
\node[sqr,draw, below = 0.55 of a15] (s15) {S};


\draw[line] (a0) -- (s0);
\draw[line] (s0) -- (b0);
\draw[line] (a1) -- (s1);
\draw[line] (s1) -- (b1);
\draw[line] (a2) -- (s2);
\draw[line] (s2) -- (b2);
\draw[line] (a13) -- (s13);
\draw[line] (s13) -- (b13);
\draw[line] (a14) -- (s14);
\draw[line] (s14) -- (b14);
\draw[line] (a15) -- (s15);
\draw[line] (s15) -- (b15);

\path[every node/.style={font=\sffamily\small}]
(l) edge [line,out=60,in=110](s0)
(b0) edge [line,out=60,in=110](s1)
(b1) edge [line,out=60,in=110](s2)
(b33) edge [line,out=60,in=110](s13)
(b13) edge [line,out=60,in=110](s14)
(b14) edge [line,out=60,in=110](s15);
\end{tikzpicture}
\caption{\texttt{Sbox from Left}}\label{lsbox}
\end{center}
\vspace{-.7 cm}
\end{figure}
In the first layer we can select plaintext $m$ and $m'$ such that the difference $\triangle m=m\oplus m'$ has only last 4 bit elements nonzero, say, $\triangle m=0\textsf{x} 00\cdots0\triangle x_{15}$. Let the output difference of $\eL_{(Q,\ast)}$ function be $0\textsf{x}00\cdots 0\triangle y_{15}$. If we choose the difference pair $(\triangle x, \triangle y)$ such that in $\triangle y$  either there are all four bits $`1'$ or only two  $`1'$. With such $\triangle y$, the diffusion layer function will yield the difference of the form $0\textsf{x}00\cdots 0\triangle z_{15}$. In all other cases, the output difference will be of the form
$0\textsf{x}ff\cdots f\triangle z_{15}$. In the next round, the difference  passes through the key xor layer without getting change. Now, as the $\eR$ function (or say Sbox) layer applies from the right, the input difference of first Sbox will be of the form $(0\textsf{x}0,0\textsf{x}\triangle z)$, which produces a nonzero difference, say $\triangle u$. The input difference for the next Sbox will be of the form $(0\textsf{x}\triangle u, 0\textsf{x}0)$.
Therefore, the second Sbox also produces a nonzero difference output. Thus, all the Sboxes get active in the second round. The description presented above is visualized in Figure \ref{dc}.
%

\begin{figure}[!htb]
\begin{center}
\begin{scriptsize}
\begin{tikzpicture}[node distance=0.1cm]

\node (ll) {};
\node[right = 0.7 of ll]  (a0) {$0$};
\node[right = 0.7 of a0] (a1) {$0$};
\node[right = 0.7 of a1] (a22) {};
\node[right = 0.15 of a22] (2a) {$\cdots$};
\node[right = 0.15 of 2a] (a33) {};
\node[right = 0.7 of a33] (a14) {$0$};
\node[right = 0.7 of a14] (a15) {$\bigtriangleup x_{15}$};

\node[fit=(a0)(a1)(a22)(2a)(a33)(a14)(a15)](aa){};

\node[above = 1 of a0]  (aa0) {$0$};
\node[above = 1  of a1] (aa1) {$0$};
\node[above = 1 of a22] (aa22) {};
\node[above = 1  of 2a] (2aa) {$\cdots$};
\node[above = 1  of a33] (aa33) {};
\node[above = 1  of a14] (aa14) {$0$};
\node[above = 0.95  of a15] (aa15) {$\bigtriangleup x_{15}$};

\node[below = 1.2 of a0]  (b0) {$0$};
\node[below = 1.2  of a1] (b1) {$0$};
\node[below = 1.2 of a22] (b22) {};
\node[below = 1.3  of 2a] (2b) {$\cdots$};
\node[below = 1.4  of a33] (b33) {};
\node[below = 1.2  of a14] (b14) {$0$};
\node[below = 1.15  of a15] (b15) {$\bigtriangleup y_{15}$};
\node[ fit=(b0)(b1)(b22)(2b)(b33)(b14)(b15)](bb){};
\node[right = 0.8 of b15](l2){};
\node[below = 1.25 of ll](l){$l$};

\node[sqr3,draw, below = 0.35 of a0] (s0) {S};
\node[sqr3,draw, below = 0.35 of a1] (s1) {S};
\node[below = 0.4  of a22] (2s) {};
\node[below = 0.45  of 2a] (2ss) {$\cdots $};
\node[sqr3,draw, below = 0.35 of a14] (s14) {S};
\node[sqrb3,draw, below = 0.30 of a15] (s15) {\textbf{S}};

\node[below = 0.7 of b0]  (d0) {$0$};
\node[below = 0.7 of b1] (d1) {$0$};
\node[below = 0.7 of b22] (d22) {};

\node[below = 0.75 of 2b] (2d) {$\cdots$};

\node[below = 0.7 of b33] (d33) {};
\node[below = 0.7 of b14] (d14) {$0$};
\node[below = 0.63 of b15] (d15) {$\bigtriangleup z_{15} $};
\node[ fit=(d0)(d1)(d22)(2d)(d33)(d14)(d15)](dd){};
\node[below = 0.77 of l2](l3){$0$};
\node[rec3, below = 0.2 of dd](key1){kXor};
\node[rec3, above = 2.6 of dd](key0){kXor};
\node[ above = 0.12 of $(d0.north)!0.5!(d1.north)$] (eo1) {$\oplus$};
\node[ above = 0.15 of $(d14.north)!0.5!(d15.north)$] (eo4) {$\oplus$};
\node[ above = 0.17 of $(d15.north)!0.5!(l3.north)$] (eo5) {$\oplus$};
%
\draw (l3) -- (eo5);
\draw (d15) -- (eo4);
\draw (d1) -- (eo1);
\draw[line] (eo5) -- (b15);
\draw[line] (eo4) -- (b14);
\draw[line] (eo1) -- (b0);
\draw[line] (b0) -- (d0);
\draw[line] (b1) -- (d1);
\draw[line] (b14) -- (d14);
\draw[line] (b15) -- (d15);

\draw[line] (a0) -- (s0);
\draw[line] (s0) -- (b0);
\draw[line] (a1) -- (s1);
\draw[line] (s1) -- (b1);
\draw[line] (a14) -- (s14);
\draw[line] (s14) -- (b14);

\draw[line] (a15) -- (s15);
\draw[line] (s15) -- (b15);

\path[every node/.style={font=\sffamily\small}]
(l) edge [line,out=45,in=130](s0)
(b0) edge [line,out=45,in=130](s1)
(b1) edge [line,out=45,in=130](2s)
(b33) edge [line,out=45,in=130](s14)
(b14) edge [line,out=40,in=130](s15);

\node[below = 1 of d0]  (a0_2) {$\bigtriangleup u_{1},0$};
\node[below = 1 of d1] (a1_2) {$\bigtriangleup u_{2},0$};
\node[below = 1 of d22] (a22_2) {};
\node[below = 1.1 of 2d] (2a_2) {$\cdots$};
\node[below = 1 of d33] (a33_2) {};
\node[below = 1 of d14] (a14_2) {$\bigtriangleup u_{15},0$};
\node[below = 0.95 of d15] (a15_2) {$0,\bigtriangleup z_{15}$};

\node[fit=(a0_2)(a1_2)(a22_2)(2a_2)(a33_2)(a14_2)(a15_2)](aa_2){};

\node[below = 1.2 of a0_2]  (b0_2) {$\bigtriangleup u_{0}$};
\node[below = 1.2  of a1_2] (b1_2) {$\bigtriangleup u_{1}$};
\node[below = 1.9 of a22_2] (b22_2) {};
\node[below = 1.3  of 2a_2] (2b_2) {$\cdots$};
\node[below = 1.2  of a33_2] (b33_2) {};
\node[below = 1.2  of a14_2] (b14_2) {$\bigtriangleup u_{14}$};
\node[below = 1.15  of a15_2] (b15_2) {$\bigtriangleup u_{15}$};
\node[ fit=(b0_2)(b1_2)(b22_2)(2b_2)(b33_2)(b14_2)(b15_2)](bb_2){};
\node[right = 0.8 of b15_2](l2_2){};
\node[below = 2.55 of l3](l_2){$l^{'}$};

\node[sqrb3,draw, below = 0.35 of a0_2] (s0_2) {\textbf{S}};
\node[sqrb3,draw, below = 0.35 of a1_2] (s1_2) {\textbf{S}};
\node[below = 1 of a33_2] (2s_2) {};
\node[below = 0.5  of 2a_2] (2ss_2) {$\cdots $};
\node[sqrb3,draw, below = 0.35 of a14_2] (s14_2) {\textbf{S}};
\node[sqrb3,draw, below = 0.30 of a15_2] (s15_2) {\textbf{S}};

%
%


\draw[line] (a0_2) -- (s0_2);
\draw[line] (s0_2) -- (b0_2);
\draw[line] (a1_2) -- (s1_2);
\draw[line] (s1_2) -- (b1_2);
\draw[line] (a14_2) -- (s14_2);
\draw[line] (s14_2) -- (b14_2);

\draw[line] (a15_2) -- (s15_2);
\draw[line] (s15_2) -- (b15_2);

\path[every node/.style={font=\sffamily\small}]
(l_2) edge [line,out=130,in=50](s15_2)
(b15_2) edge [line,out=130,in=50](s14_2)
(b14_2) edge [line,out=130,in=50](2s_2)
(b22_2) edge [line,out=130,in=50](s1_2)
(b1_2) edge [line,out=135,in=50](s0_2);

\end{tikzpicture}
\caption{\texttt{Differential Cryptanalysis}}\label{dc}
\end{scriptsize}
\end{center}
\vspace{-4 mm}
\end{figure}
Hence, the (standard) differential cryptanalysis techniques do not provides a distinguisher. 
{ \begin{remark}
We have also carried out linear and differential cryptanalysis of the design on the decryption algorithm. The decryption algorithm uses $\dL$ and $\dR$ operations over the conjugate of the quasigroup, $(Q,\backslash _\ast)$.  We have found that the number of active Sboxes increases considerably, raising the data complexity more than the maximum possible pairs. 
\end{remark} }
\subsection{Algebraic Cryptanalysis}
 Any cryptographic design must be resistant to algebraic cryptanalysis \cite {bard2009algebraic}. The binary operation $\ast$ of a quasigroup  of order $2^d$ can be represented  by a vector of Boolean polynomials.  We use a (polynomially complete) quasigroup $(Q,\ast)$ of order $2^4$.   So, the elements $x \in Q$ can be represented as $(x_0,x_1,x_2,x_3)$, $x_i\in \{0,1\}$. There exists a set of 4 Boolean polynomials $f_0,f_1,f_2,f_3$ such that 
\begin{equation}\label{quasivbf}
x\ast x' = \big(f_0(x,x'), f_1(x,x'), f_2(x,x'), f_3(x,x')\big), \quad \forall x,x'\in Q
\end{equation} 
The Boolean polynomials $f_0,f_1,f_2,f_3$ corresponding to the quasigroup used in the design are given in the Appendix. The coordinate functions $f_i's$ are nonlinear with algebraic degree $6$. The degree of all nonzero linear combinations of $f_0,f_1,f_2$ and $f_3$ remain the same, that is $6$.
So, the algebraic degree of the vector Boolean representation of $Q$ is $6$.
 The algebraic degree of all component functions corresponding to the coordinate  Boolean functions of the conjugate quasigroup $(Q,\backslash_{\ast})$ is also $6$. 
\par
Let $64$-bit input block of plaintext be $x_0,\ldots,x_{63}$. If one denotes the first round key variables by $k^1_0,\ldots,k^1_{63}$,  the linear Boolean  equations of the key xor layer are  \vspace{-4 mm}
  \begin{eqnarray*}
 y_i=k^1_i+x_i, ~i=0,\ldots,63,
\end{eqnarray*}
where $y_i's$ are the output bit variables. Further, one can express $\eL_{(Q,\ast)}$ string transformation  using the Boolean functions $f_0,f_1,f_2$ and $f_3$ corresponding to $(Q,\ast)$ in the following way.
As the leader for $\eL$ in the first round is $l=(k^1_0,k^1_1,k^1_2,k^1_3)$, the output bits variables $z_0,z_1,\ldots,z_{63}$ of $\eL$ map:
\begin{eqnarray*}
z_0&=&f_0(k^1_0,k^1_1,k^1_2,k^1_3,y_0,y_1,y_2,y_3)\\
z_1&=&f_1(k^1_0,k^1_1,k^1_2,k^1_3,y_0,y_1,y_2,y_3)\\
z_2&=&f_2(k^1_0,k^1_1,k^1_2,k^1_3,y_0,y_1,y_2,y_3)\\
z_3&=&f_3(k^1_0,k^1_1,k^1_2,k^1_3,y_0,y_1,y_2,y_3)\\
z_4&=&f_0(z_0,z_1,z_2,z_3,y_4,y_5,y_6,y_7), \text{ and so on}
\end{eqnarray*}
\\ \noindent
Let us consider  the binary variables of the output of diffusion function $\eR_{(\mathbb{F}_2,\oplus),0}$ are $u_0,u_1,\ldots,u_{63}$, then
\begin{eqnarray*}
u_{63} = z_{63}+0,\quad u_{i} = z_{i}+u_{i+1}, \quad i=62\ldots 0
\end{eqnarray*}

The above equations for $y_i,z_i$ and $u_i$ in terms of $x_i$ and $k^1_i$ provide a (nonlinear multivariate) algebraic system over $\mathbb{F}_2$ for one round. In the same manner, we can obtain the equations for the other rounds in terms of intermediate variables and then finally equations for ciphertext bits, say $c_0,\ldots,c_{63}$.  The (complete) algebraic system consists of $2^{10}$ six degree equations in approx $2^{11}$ binary variables. In the counting, we have considered only those variables which cannot be eliminated using linear equations of the system.

For a known pair of plaintext-ciphertext, one can try to solve the system for the key variables. But the number of variables and degree of the system makes very difficult  to solve the above system and recover some secret information about the design. As solving  nonlinear multivariate system for a sufficiently large system in enough number of variables is NP-complete. 
 We modeled reduced design of only two rounds in polynomial equations over $\mathbb{F}_2$. The polynomial system is expressed in $256$ variables. We perform the experiments to solve the system by using Gr\"obner basis's  implementations \texttt{std} and \texttt{slimgb} of \textsf{Singular}\cite{singular}.
We use a server: Intel(R) Core(TM) $i7-8700$
cpu, $3.20$ GHz, $6$ cores, $12$ threads, $32$ gigabyte RAM with a Debian based linux operating system. We killed the \texttt{slimgb}  computation as the memory requirement reached to $25$ gigabytes. The \texttt{std} computations ran for more than $3$ hours. There were $455161$ ``s-polynomials'' to be reduced at the time when computation got killed.
So, by these experiments, we strongly believe  that our design is resistant to the algebraic attacks.
\section{Randomness Testing of the Design}\label{s4}
In this section, we evaluate the randomness of our design algorithm using {NIST} statistical test suite { \cite{10.5555/2206233}} and compare the experimental results with AES-128.
\subsection*{Experimental setup} We apply the NIST test suite on $64$ binary sequences of length $2^{20}$ generated using $64$ random $128$-bit keys in CBC, CFB, OFB and CTR mode of operations of our design and AES-128, and then compare their \emph{p-values}. The tests have been carried out with the significance level of $99\%$. Table \ref{t1} and \ref{t2} provide the \emph{p-values} of the experiment conducted with plaintext of all zeros and all ones. The abbreviation used for the NIST tests in the following tables are given in Appendix. The column \texttt{INRU} in the table represents data corresponding to our design.  

\begin{table}[h!]
\begin{center}
\begin{scriptsize}
\begin{tabular}{|c|c|c|c|c|c|c|c|c|}
\hline 
& \multicolumn{2}{|c|}{CBC} &  \multicolumn{2}{|c|}{CTR} & \multicolumn{2}{|c|}{CFB} & \multicolumn{2}{|c|}{OFB} \\ \hline
Test	&	AES	&	\texttt{INRU}	&	AES	&\texttt{INRU}	&	AES	&	\texttt{INRU}	&	AES	&	\texttt{INRU}	\\ \hline
\textbf{AE}	&	0.4965	&	\textbf{0.5464}	&	0.4440	&	\textbf{0.4585}	&	0.4965	&	0.4205	&	0.4579	&	\textbf{0.5300}	\\ \hline
\textbf{BF}	&	0.5069	&	\textbf{0.5281}	&	0.4902	&	\textbf{0.5512}	&	0.5069	&	\textbf{0.5260}	&	0.4971	&	\textbf{0.5230}	\\ \hline
\textbf{CSF}	&	0.5320	&	0.4788	&	0.4892	&	0.4873	&	0.5320	&	0.5195	&	0.5273	&	\textbf{0.5275}	\\ \hline
\textbf{CSB}	&	0.5359	&	0.4804	&	0.5314	&	0.5050	&	0.5359	&	0.4828	&	0.5474	&	0.5261	\\ \hline
\textbf{DFT}	&	0.5076	&	0.4909	&	0.4838	&	\textbf{0.4868}	&	0.5076	&	0.5030	&	0.4808	&	\textbf{0.5184}	\\ \hline
\textbf{LC}	&	0.4944	&	0.4675	&	0.4805	&	\textbf{0.4992}	&	0.4944	&	0.4859	&	0.4990	&	\textbf{0.5302}	\\ \hline
\textbf{Freq}	&	0.5529	&	0.4775	&	0.5364	&	0.5017	&	0.5529	&	0.4833	&	0.5037	&	\textbf{0.5533}	\\ \hline
\textbf{LRO}	&	0.4781	&	0.4747	&	0.4953	&	\textbf{0.4953}	&	0.4781	&	0.4753	&	0.5043	&	0.4754	\\ \hline
\textbf{NOTM}	&	0.5019	&	\textbf{0.5020}	&	0.5010	&	0.4992	&	0.5019	&	0.5003	&	0.4979	&	\textbf{0.5051}	\\ \hline
\textbf{OTM}	&	0.4574	&	\textbf{0.5714}	&	0.4544	&	\textbf{0.4798}	&	0.4574	&	\textbf{0.4987}	&	0.5125	&	\textbf{0.5134}	\\ \hline
\textbf{Rank}	&	0.4528	&	\textbf{0.5485}	&	0.5901	&	0.4635	&	0.4528	&	\textbf{0.5237}	&	0.4939	&	\textbf{0.5158}	\\ \hline
\textbf{Run}	&	0.4380	&	0.4260	&	0.4873	&	0.4513	&	0.4380	&	\textbf{0.5093}	&	0.4431	&	\textbf{0.4601}	\\ \hline
\textbf{Srl}	&	0.5208	&	0.4948	&	0.4907	&	\textbf{0.5225}	&	0.5208	&	0.4569	&	0.5050	&	\textbf{0.5253}	\\ \hline
\textbf{Univ}	&	0.5247	&	0.5174	&	0.4696	&	\textbf{0.5565}	&	0.5247	&	0.5041	&	0.4947	&	\textbf{0.5114}	\\ \hline
\end{tabular}

\caption{All Zeroes Input}\label{t1}
\end{scriptsize}
\end{center}
\end{table}

\begin{table}[h!]
\begin{center}

\begin{scriptsize}
\begin{tabular}{|c|c|c|c|c|c|c|c|c|}
\hline 
& \multicolumn{2}{|c|}{CBC} &  \multicolumn{2}{|c|}{CTR} & \multicolumn{2}{|c|}{CFB} & \multicolumn{2}{|c|}{OFB} \\ \hline
Test	&	AES	&	\texttt{INRU}	&	AES	&	\texttt{INRU}	&	AES	&	\texttt{INRU}	&	AES	&	\texttt{INRU}	\\ \hline
\textbf{AE}	&	0.5059	&	0.4997	&	0.4440	&	\textbf{0.4585}	&	0.4873	&	0.4782	&	0.4579	&	\textbf{0.5300}	\\ \hline
\textbf{BF}	&	0.5047	&	0.4605	&	0.4902	&	\textbf{0.5512}	&	0.5533	&	0.5505	&	0.4971	&	\textbf{0.5230}	\\ \hline
\textbf{CSF}	&	0.4966	&	0.4514	&	0.4892	&	0.4873	&	0.5388	&	0.5218	&	0.5273	&	\textbf{0.5275}	\\ \hline
\textbf{CSB}	&	0.4995	&	0.4415	&	0.5314	&	0.5050	&	0.5279	&	0.5030	&	0.5474	&	0.5261	\\ \hline
\textbf{DFT}	&	0.5643	&	0.5443	&	0.4838	&	\textbf{0.4868}	&	0.4960	&	0.4725	&	0.4808	&	\textbf{0.5184	}\\ \hline
\textbf{LC}	&	0.5189	&	0.5182	&	0.5634	&	0.5124	&	0.5099	&	\textbf{0.5324}	&	0.5200	&	0.4750	\\ \hline
\textbf{Freq}	&	0.5060	&	0.4439	&	0.5364	&	0.5017	&	0.5231	&	0.5179	&	0.5037	&	\textbf{0.5533}	\\ \hline
\textbf{LRO}	&	0.5022	&	\textbf{0.5195}	&	0.4996	&	0.4993	&	0.5331	&	\textbf{0.5361}	&	0.5418	&	0.4797	\\ \hline
\textbf{NOTM}	&	0.5034	&	0.4907	&	0.5010	&	0.4992	&	0.4995	&	\textbf{0.5008}	&	0.4979	&	\textbf{0.5051}	\\ \hline
\textbf{OTM}	&	0.4652	&	\textbf{0.5010}	&	0.4757	&	\textbf{0.4874}	&	0.4534	&	\textbf{0.5705}	&	0.5241	&	0.4461	\\ \hline
\textbf{Rank}	&	0.5129	&	0.4624	&	0.4798	&	\textbf{0.5110}	&	0.5202	&	0.4701	&	0.5001	&	0.4236	\\ \hline
\textbf{Run}	&	0.4893	&	0.4789	&	0.4873	&	0.4513	&	0.4885	&	0.4742	&	0.4431	&	\textbf{0.4601}	\\ \hline
\textbf{Srl}	&	0.5024	&	\textbf{0.5487}	&	0.4907	&	\textbf{0.5225}	&	0.4984	&	0.4616	&	0.5050	&	\textbf{0.5253}	\\ \hline
\textbf{Univ}	&	0.5253	&	0.4494	&	0.4696	&	\textbf{0.5565}	&	0.5206	&	0.5044	&	0.4947	&	\textbf{0.5114}	\\ \hline
\end{tabular}
\caption{All Ones Input}\label{t2}
\end{scriptsize}
\end{center}
\end{table}
The generated sequences pass all NIST tests successfully. The table entries in bold font present better \emph{p-values} than the AES. We conclude that the randomness of our algorithm is equivalent to AES-128. The randomness testing of the design is also carried out on approx $100$ other  combinations of plaintexts and keys of low and high densities, and we found that all the NIST tests pass successfully for each combination.   
\subsection*{Avalanche Analysis of Plaintext}
 We performed avalanche analysis \cite {CASTRO20051} on our design  to measure the diffusion produced by the encryption algorithm. The avalanche effect was analysed by changing each bit of $10000$ random inputs with $6$ different random keys. We compute the avalanche percentage by xoring ciphertext of original plaintext and ciphertext of perturbed plaintext at each bit. The following table presents the avalanche range of $95\%, 98\%$ and $99\%$ of the experiments. 
\begin{table}[H]
\begin{center}\vspace{-4 mm}
\begin{tabular}{|c|c|c|}
\hline
$95\%$&$98\%$&$99\%$\\
\hline
 $(48.49,~51.54)$ &    $(48.19,~ 51.82)$ &     $(48.00,~52.00)$\\
 \hline
\end{tabular}
\end{center}
\label{avlp}
\end{table}
\vspace{-4 mm}
We also measured the dependence of all plaintext bits on each ciphertext bit (\emph{strict avalanche analysis}) and tabulated the result below:
\begin{table}[H]
\begin{center}\vspace{-4 mm}
\begin{tabular}{|c|c|c|}
\hline
$95\%$&$98\%$&$99\%$\\
\hline
 $(48.18,~   51.63)$ &    $(47.82,~   51.99)$ &     $(47.50,~   52.18)$\\
 \hline
\end{tabular}
\end{center}
\label{strrlrl}
\end{table} 
\vspace{-4 mm}
%

\subsection*{Key Scheduling Attacks} As, the round keys are being generated using string transformations $\eL$ and $\eR$ which are nonlinear functions and carrying dependence of all preceding master key bits in the forward direction, see Figures { \ref{figeleft}} and { \ref{figeright}}. Each bit of round keys depends non-linearly on all bits of the master key since the quasigroup  operation is nonlinear of degree $6$.
So, we believe that  such a degree of non-linearity and dependence of each round key bit on all master key bits  are enough to resist key scheduling attacks, namely related key attacks, slide attack etc.{ \cite{Biryukovrelated2011, Biryukovslide2011}}. 
\section{Hardware Performance Analysis}
We have estimated the gate equivalent (GE) count of the Encryption module of INRU. Here we assumed that round keys are either stored in external memory or provided as inputs at appropriate clock cycles. This is justified since the key expansion is performed just once for the master key. For checking the hardware complexity, the different components of round function of INRU are implemented using Verilog hardware description
language. The ASIC based hardware simulation of the design has been carried out by
synthesizing via Mentor Graphics Leonardo Spectrum with the $90$ nm ($0.09$ $\mu m$) technology
node. Our circuit processed one round of encryption per $12$ clock cycles. The simulation result
shows that the total size corresponds to $2125$ GE. The operating clock frequency is set to be
$500$ MHz. So the data throughput is about $166.6$ Mbps.

\section{Conclusion}\label{s5}

In this paper, we proposed quasigroup based block cipher INRU with 64-bit block length and 128-bit key
length. It is a actually a lightweight block cipher. It can be implemented in resource
constrained IOT devices.  We have carried out the standard linear and differential cryptanalysis and shown that these techniques do not reduce security strength. We have also tried to solve algebraic systems over $\mathbb{F}_2$ for 2 rounds of the design for known plaintext-ciphertext pairs using Gr\"obner bases engines \texttt{std} and \texttt{slimgb} of {\sc Singular}. Based on the experiments we believe that the design is robust against the algebraic attacks. We also carried out the statistical analysis to check the randomness of the algorithm using NIST test suite and conclude that the randomizing ability of the algorithm is equivalent to that of AES-128.
In future, we will work towards optimization of the hardware implementation with respect to area and throughput.
%

\section{Acknowledgement} 
We are thankful to  Ms. U. Jeya Santhi, Director SAG, DRDO for her support and encouragement to carry out this work. We thank to Prof. Bimal Kumar Roy, ISI Kolkata for valuable comments during several discussions.
We would like to thank Dr. A.E. Pankratiev  and Dr. A.V. Galatenko, Moscow State University for carefully reading the manuscript and giving valuable suggestions regarding the security and performance analysis of the design. We also thank Dr. Swagata Mandal, Jalpaiguri Government Engineering College, for the simulation results of the hardware performance analysis. 

\bibliography{QuasigroupBasedBlockCipher}
\bibliographystyle{plain}
\section{Appendix}\label{append}
The Boolean polynomials $f_0,f_1,f_2,f_3$ of the quasigroup $(Q,\ast)$ used in the design:\newline
\begin{scriptsize}
$
f_0(x_0,x_1,x_2,x_3,x'_0,x'_1,x'_2,x'_3)=x_0*x_1*x_2*x'_0*x'_1*x'_2+x_1*x_2*x_3*x'_0*x'_1*x'_2+x_1*x_2*x_3*x'_0*x'_1*x'_3+x_0*x_2*x_3*x'_0*x'_2*x'_3+x_1*x_2*x_3*x'_0*x'_2*x'_3+x_1*x_2*x_3*x'_1*x'_2*x'_3+x_0*x_1*x_3*x'_0*x'_1+x_0*x_2*x_3*x'_0*x'_1+x_1*x_2*x_3*x'_0*x'_1+x_0*x_2*x_3*x'_1*x'_2+x_0*x_1*x'_0*x'_1*x'_2+x_0*x_2*x'_0*x'_1*x'_2+x_1*x_2*x'_0*x'_1*x'_2+x_0*x_3*x'_0*x'_1*x'_2+x_1*x_3*x'_0*x'_1*x'_2+x_2*x_3*x'_0*x'_1*x'_2+x_0*x_1*x_2*x'_0*x'_3+x_1*x_2*x_3*x'_0*x'_3+x_0*x_1*x_2*x'_1*x'_3+x_0*x_2*x_3*x'_1*x'_3+x_0*x_1*x'_0*x'_1*x'_3+x_2*x_3*x'_0*x'_1*x'_3+x_1*x_2*x_3*x'_2*x'_3+x_1*x_2*x'_0*x'_2*x'_3+x_0*x_3*x'_0*x'_2*x'_3+x_1*x_3*x'_0*x'_2*x'_3+x_0*x_1*x'_1*x'_2*x'_3+x_0*x_2*x'_1*x'_2*x'_3+x_1*x_2*x'_1*x'_2*x'_3+x_2*x_3*x'_1*x'_2*x'_3+x_0*x_1*x_3*x'_0+x_0*x_2*x_3*x'_0+x_1*x_2*x_3*x'_0+x_0*x_1*x_3*x'_1+x_1*x_3*x'_0*x'_1+x_2*x_3*x'_0*x'_1+x_0*x_1*x_3*x'_2+x_0*x_2*x_3*x'_2+x_0*x_1*x'_1*x'_2+x_0*x_2*x'_1*x'_2+x_0*x_3*x'_1*x'_2+x_1*x_3*x'_1*x'_2+x_2*x_3*x'_1*x'_2+x_0*x'_0*x'_1*x'_2+x_3*x'_0*x'_1*x'_2+x_0*x_1*x_2*x'_3+x_0*x_1*x'_0*x'_3+x_0*x_2*x'_0*x'_3+x_1*x_2*x'_0*x'_3+x_0*x_3*x'_0*x'_3+x_1*x_3*x'_0*x'_3+x_2*x_3*x'_0*x'_3+x_0*x_1*x'_1*x'_3+x_2*x_3*x'_1*x'_3+x_0*x'_0*x'_1*x'_3+x_0*x_1*x'_2*x'_3+x_2*x_3*x'_2*x'_3+x_0*x'_0*x'_2*x'_3+x_3*x'_0*x'_2*x'_3+x_1*x'_1*x'_2*x'_3+x_2*x'_1*x'_2*x'_3+x_0*x_1*x_2+x_1*x_2*x_3+x_0*x_2*x'_0+x_1*x_2*x'_0+x_1*x_3*x'_0+x_2*x_3*x'_0+x_0*x_2*x'_1+x_1*x_2*x'_1+x_0*x_3*x'_1+x_2*x_3*x'_1+x_0*x'_0*x'_1+x_1*x'_0*x'_1+x_0*x_1*x'_2+x_0*x_2*x'_2+x_1*x_2*x'_2+x_1*x_3*x'_2+x_0*x'_1*x'_2+x_1*x'_1*x'_2+x_2*x'_1*x'_2+x_3*x'_1*x'_2+x_0*x_2*x'_3+x_0*x_3*x'_3+x_1*x_3*x'_3+x_0*x'_0*x'_3+x_3*x'_0*x'_3+x_0*x'_1*x'_3+x_1*x'_1*x'_3+x_2*x'_1*x'_3+x_3*x'_1*x'_3+x_0*x'_2*x'_3+x'_1*x'_2*x'_3+x_0*x_1+x_0*x_2+x_1*x_2+x_0*x_3+x_1*x_3+x_2*x'_0+x_0*x'_1+x_2*x'_1+x'_0*x'_1+x_0*x'_2+x_2*x'_2+x'_1*x'_2+x'_0*x'_3+x'_1*x'_3+x_1+x_2+x'_1+x'_3+1,
\\[2mm]
f_1(x_0,x_1,x_2,x_3,x'_0,x'_1,x'_2,x'_3)=x_1*x_2*x_3*x'_0*x'_1*x'_3+x_0*x_1*x_3*x'_0*x'_2*x'_3+x_0*x_1*x_2*x'_1*x'_2*x'_3+x_1*x_2*x_3*x'_1*x'_2*x'_3+x_0*x_1*x_3*x'_0*x'_1+x_0*x_2*x_3*x'_0*x'_1+x_0*x_1*x_3*x'_0*x'_2+x_0*x_2*x_3*x'_0*x'_2+x_1*x_2*x_3*x'_0*x'_2+x_0*x_1*x_2*x'_1*x'_2+x_0*x_1*x_3*x'_1*x'_2+x_0*x_1*x'_0*x'_1*x'_2+x_0*x_2*x'_0*x'_1*x'_2+x_1*x_2*x'_0*x'_1*x'_2+x_0*x_3*x'_0*x'_1*x'_2+x_2*x_3*x'_0*x'_1*x'_2+x_0*x_1*x_2*x'_0*x'_3+x_0*x_2*x_3*x'_0*x'_3+x_0*x_1*x_2*x'_1*x'_3+x_0*x_2*x_3*x'_1*x'_3+x_0*x_1*x'_0*x'_1*x'_3+x_0*x_1*x_2*x'_2*x'_3+x_0*x_2*x_3*x'_2*x'_3+x_1*x_2*x_3*x'_2*x'_3+x_0*x_2*x'_0*x'_2*x'_3+x_1*x_2*x'_0*x'_2*x'_3+x_0*x_3*x'_1*x'_2*x'_3+x_1*x_3*x'_1*x'_2*x'_3+x_2*x_3*x'_1*x'_2*x'_3+x_0*x_1*x_2*x'_0+x_1*x_2*x_3*x'_0+x_0*x_1*x_3*x'_1+x_0*x_2*x_3*x'_1+x_1*x_2*x_3*x'_1+x_0*x_1*x'_0*x'_1+x_1*x_2*x'_0*x'_1+x_1*x_3*x'_0*x'_1+x_0*x_2*x_3*x'_2+x_1*x_2*x_3*x'_2+x_1*x_3*x'_0*x'_2+x_2*x_3*x'_0*x'_2+x_0*x_1*x'_1*x'_2+x_0*x_2*x'_1*x'_2+x_0*x_3*x'_1*x'_2+x_1*x_3*x'_1*x'_2+x_0*x_1*x_3*x'_3+x_0*x_2*x_3*x'_3+x_0*x_2*x'_0*x'_3+x_1*x_2*x'_0*x'_3+x_0*x_3*x'_0*x'_3+x_0*x_1*x'_1*x'_3+x_2*x_3*x'_1*x'_3+x_2*x_3*x'_2*x'_3+x_1*x'_0*x'_2*x'_3+x_2*x'_0*x'_2*x'_3+x_1*x'_1*x'_2*x'_3+x_3*x'_1*x'_2*x'_3+x_0*x_1*x_2+x_0*x_2*x_3+x_1*x_2*x'_0+x_0*x_3*x'_0+x_1*x_3*x'_0+x_0*x_1*x'_1+x_0*x_2*x'_1+x_1*x_2*x'_1+x_0*x_3*x'_1+x_1*x_3*x'_1+x_1*x'_0*x'_1+x_3*x'_0*x'_1+x_0*x_1*x'_2+x_0*x_3*x'_2+x_1*x_3*x'_2+x_0*x'_0*x'_2+x_1*x'_0*x'_2+x_0*x'_1*x'_2+x_2*x'_1*x'_2+x_0*x_2*x'_3+x_0*x_3*x'_3+x_1*x_3*x'_3+x_2*x_3*x'_3+x_0*x'_0*x'_3+x_1*x'_0*x'_3+x_2*x'_0*x'_3+x_3*x'_0*x'_3+x_0*x'_1*x'_3+x_1*x'_1*x'_3+x_2*x'_1*x'_3+x_3*x'_1*x'_3+x'_0*x'_1*x'_3+x_0*x'_2*x'_3+x_1*x'_2*x'_3+x_2*x'_2*x'_3+x_3*x'_2*x'_3+x'_0*x'_2*x'_3+x_0*x_2+x_0*x_3+x_1*x_3+x_0*x'_0+x_1*x'_0+x_0*x'_1+x_1*x'_1+x'_0*x'_1+x_0*x'_2+x_2*x'_2+x'_1*x'_2+x_1*x'_3+x_3*x'_3+x'_0*x'_3+x_0+x_2+x_3+x'_0+x'_1,
\\[2mm]
f_2(x_0,x_1,x_2,x_3,x'_0,x'_1,x'_2,x'_3)=x_1*x_2*x_3*x'_0*x'_1*x'_3+x_1*x_2*x_3*x'_0*x'_2*x'_3+x_0*x_1*x_2*x'_1*x'_2*x'_3+x_0*x_2*x_3*x'_1*x'_2*x'_3+x_1*x_2*x_3*x'_1*x'_2*x'_3+x_0*x_1*x_2*x'_0*x'_1+x_0*x_2*x_3*x'_0*x'_1+x_1*x_2*x_3*x'_0*x'_1+x_0*x_1*x_3*x'_0*x'_2+x_0*x_2*x_3*x'_0*x'_2+x_0*x_1*x_2*x'_1*x'_2+x_1*x_2*x'_0*x'_1*x'_2+x_1*x_3*x'_0*x'_1*x'_2+x_0*x_1*x_3*x'_0*x'_3+x_0*x_2*x_3*x'_0*x'_3+x_0*x_1*x_2*x'_1*x'_3+x_1*x_2*x_3*x'_1*x'_3+x_0*x_1*x'_0*x'_1*x'_3+x_0*x_3*x'_0*x'_1*x'_3+x_1*x_3*x'_0*x'_1*x'_3+x_2*x_3*x'_0*x'_1*x'_3+x_0*x_1*x_2*x'_2*x'_3+x_0*x_2*x_3*x'_2*x'_3+x_0*x_1*x'_0*x'_2*x'_3+x_0*x_3*x'_0*x'_2*x'_3+x_1*x_3*x'_0*x'_2*x'_3+x_2*x_3*x'_0*x'_2*x'_3+x_0*x_1*x'_1*x'_2*x'_3+x_0*x_2*x'_1*x'_2*x'_3+x_1*x_2*x'_1*x'_2*x'_3+x_0*x_1*x_2*x'_0+x_0*x_2*x_3*x'_0+x_0*x_1*x_3*x'_1+x_0*x_2*x_3*x'_1+x_0*x_1*x'_0*x'_1+x_0*x_2*x'_0*x'_1+x_1*x_2*x'_0*x'_1+x_0*x_1*x_2*x'_2+x_0*x_1*x_3*x'_2+x_0*x_2*x_3*x'_2+x_1*x_2*x_3*x'_2+x_0*x_1*x'_0*x'_2+x_0*x_2*x'_0*x'_2+x_1*x_2*x'_0*x'_2+x_1*x_3*x'_0*x'_2+x_0*x_1*x'_1*x'_2+x_0*x_2*x'_1*x'_2+x_0*x'_0*x'_1*x'_2+x_1*x'_0*x'_1*x'_2+x_2*x'_0*x'_1*x'_2+x_0*x_1*x_2*x'_3+x_0*x_1*x_3*x'_3+x_0*x_1*x'_0*x'_3+x_0*x_2*x'_0*x'_3+x_0*x_2*x'_1*x'_3+x_0*x_3*x'_1*x'_3+x_1*x_3*x'_1*x'_3+x_2*x_3*x'_1*x'_3+x_2*x'_0*x'_1*x'_3+x_3*x'_0*x'_1*x'_3+x_0*x_1*x'_2*x'_3+x_0*x_2*x'_2*x'_3+x_1*x_2*x'_2*x'_3+x_0*x_3*x'_2*x'_3+x_1*x_3*x'_2*x'_3+x_2*x_3*x'_2*x'_3+x_2*x'_0*x'_2*x'_3+x_3*x'_0*x'_2*x'_3+x_0*x'_1*x'_2*x'_3+x_2*x'_1*x'_2*x'_3+x_3*x'_1*x'_2*x'_3+x_0*x_1*x_2+x_1*x_2*x_3+x_0*x_1*x'_0+x_0*x_2*x'_0+x_1*x_2*x'_0+x_0*x_3*x'_0+x_1*x_3*x'_0+x_2*x_3*x'_0+x_0*x_1*x'_1+x_0*x_2*x'_1+x_0*x'_0*x'_1+x_2*x'_0*x'_1+x_3*x'_0*x'_1+x_1*x_2*x'_2+x_0*x_3*x'_2+x_2*x_3*x'_2+x_0*x'_0*x'_2+x_2*x'_0*x'_2+x_0*x'_1*x'_2+x_3*x'_1*x'_2+x_0*x_3*x'_3+x_2*x_3*x'_3+x_1*x'_0*x'_3+x_0*x'_1*x'_3+x_2*x'_1*x'_3+x_0*x'_2*x'_3+x_2*x'_2*x'_3+x'_1*x'_2*x'_3+x_0*x_1+x_0*x_2+x_1*x_2+x_0*x_3+x_1*x_3+x_2*x_3+x_0*x'_0+x_2*x'_0+x_1*x'_1+x'_0*x'_1+x_2*x'_2+x'_1*x'_2+x'_1*x'_3+x'_1+x'_2+1,
\\[2mm]
f_3(x_0,x_1,x_2,x_3,x'_0,x'_1,x'_2,x'_3)=x_0*x_2*x_3*x'_0*x'_1*x'_2+x_1*x_2*x_3*x'_0*x'_1*x'_2+x_0*x_1*x_2*x'_0*x'_1*x'_3+x_0*x_1*x_3*x'_0*x'_1*x'_3+x_0*x_2*x_3*x'_0*x'_1*x'_3+x_0*x_2*x_3*x'_0*x'_2*x'_3+x_1*x_2*x_3*x'_0*x'_2*x'_3+x_0*x_2*x_3*x'_1*x'_2*x'_3+x_0*x_1*x_2*x'_0*x'_1+x_0*x_1*x_3*x'_0*x'_1+x_1*x_2*x_3*x'_0*x'_1+x_0*x_1*x_3*x'_0*x'_2+x_0*x_2*x_3*x'_0*x'_2+x_1*x_2*x_3*x'_0*x'_2+x_0*x_1*x_2*x'_1*x'_2+x_0*x_2*x_3*x'_1*x'_2+x_1*x_2*x_3*x'_1*x'_2+x_0*x_2*x'_0*x'_1*x'_2+x_0*x_3*x'_0*x'_1*x'_2+x_0*x_1*x_2*x'_0*x'_3+x_0*x_1*x_3*x'_0*x'_3+x_0*x_2*x_3*x'_0*x'_3+x_1*x_2*x_3*x'_0*x'_3+x_0*x_2*x_3*x'_1*x'_3+x_1*x_2*x_3*x'_1*x'_3+x_1*x_2*x'_0*x'_1*x'_3+x_0*x_3*x'_0*x'_1*x'_3+x_2*x_3*x'_0*x'_1*x'_3+x_0*x_1*x_3*x'_2*x'_3+x_1*x_2*x_3*x'_2*x'_3+x_0*x_3*x'_0*x'_2*x'_3+x_2*x_3*x'_0*x'_2*x'_3+x_0*x_3*x'_1*x'_2*x'_3+x_1*x_3*x'_1*x'_2*x'_3+x_2*x_3*x'_1*x'_2*x'_3+x_1*x_2*x_3*x'_0+x_0*x_1*x_2*x'_1+x_0*x_2*x_3*x'_1+x_1*x_2*x_3*x'_1+x_0*x_1*x'_0*x'_1+x_1*x_2*x'_0*x'_1+x_1*x_3*x'_0*x'_1+x_0*x_1*x_2*x'_2+x_1*x_2*x_3*x'_2+x_1*x_3*x'_0*x'_2+x_2*x_3*x'_0*x'_2+x_0*x_1*x'_1*x'_2+x_1*x_2*x'_1*x'_2+x_1*x_3*x'_1*x'_2+x_2*x_3*x'_1*x'_2+x_0*x'_0*x'_1*x'_2+x_0*x_1*x_2*x'_3+x_0*x_1*x_3*x'_3+x_0*x_2*x_3*x'_3+x_1*x_2*x_3*x'_3+x_1*x_2*x'_0*x'_3+x_0*x_3*x'_0*x'_3+x_1*x_3*x'_0*x'_3+x_2*x_3*x'_0*x'_3+x_0*x_1*x'_1*x'_3+x_0*x_2*x'_1*x'_3+x_0*x_3*x'_1*x'_3+x_1*x_3*x'_1*x'_3+x_0*x_2*x'_2*x'_3+x_0*x_3*x'_2*x'_3+x_2*x_3*x'_2*x'_3+x_0*x'_0*x'_2*x'_3+x_1*x'_0*x'_2*x'_3+x_2*x'_0*x'_2*x'_3+x_3*x'_0*x'_2*x'_3+x_2*x'_1*x'_2*x'_3+x_0*x_1*x_2+x_0*x_2*x_3+x_1*x_2*x_3+x_0*x_1*x'_0+x_0*x_1*x'_1+x_0*x_2*x'_1+x_1*x_2*x'_1+x_1*x_3*x'_1+x_2*x'_0*x'_1+x_0*x_2*x'_2+x_1*x_2*x'_2+x_0*x_3*x'_2+x_1*x_3*x'_2+x_2*x_3*x'_2+x_0*x'_0*x'_2+x_1*x'_0*x'_2+x_2*x'_1*x'_2+x_3*x'_1*x'_2+x_0*x_2*x'_3+x_1*x_2*x'_3+x_0*x_3*x'_3+x_1*x_3*x'_3+x_0*x'_0*x'_3+x_1*x'_0*x'_3+x_3*x'_0*x'_3+x_1*x'_1*x'_3+x_3*x'_1*x'_3+x'_0*x'_1*x'_3+x_0*x'_2*x'_3+x_3*x'_2*x'_3+x_0*x_2+x_1*x_2+x_0*x_3+x_1*x'_1+x'_0*x'_1+x_0*x'_2+x_2*x'_2+x_3*x'_2+x'_0*x'_2+x'_1*x'_2+x_0*x'_3+x_2*x'_3+x'_0*x'_3+x'_2*x'_3+x_3+x'_0+x'_3
$
\end{scriptsize}
where $+$ and $*$ are the addition and multiplication operations in the field $\mathbb{F}_2$.
\begin{scriptsize}
\begin{table}[h!]
\begin{center}
\begin{tabular}{|c|c|c|c|}
\hline
\textbf{Abbreviation} & \textbf{Test} & \textbf{Abbreviation} & \textbf{Test} \\ \hline
AE & Approximate Entropy& BF&Block Frequency \\ \hline
 CSF& \makecell {Cumulative \\ Sum Forward}  & CSB& \makecell {Cumulative \\ Sum Backward }\\ \hline
 DFT& \makecell {Discrete Fourier \\ Transform } & LC& Linear Complexity \\ \hline 
Freq & Frequency &  LRO & \makecell {Longest Run of \\ Ones in a Block} \\ \hline
NOTM & \makecell {Non-overlapping \\ Template Matching} &  OTM & \makecell {Overlapping \\Template Matching} \\ \hline
Srl & Serial &  Univ & Maurer's Universal \\\hline
\end{tabular}
\caption{Abbreviations used for NIST statistical tests}
\end{center}
\end{table}
\end{scriptsize}

\end{document}